\documentclass[aps,prb,reprint,grounedaddress]{revtex4-2}

\usepackage{graphicx, graphics,color,epsfig}
\usepackage{bm}
\usepackage{dcolumn}
\usepackage{amsmath}
\usepackage{amssymb}
\usepackage{graphicx}
\usepackage{epstopdf}
\usepackage[colorlinks,
anchorcolor=blue,  
linkcolor=blue,       
citecolor=blue]{hyperref}  
\usepackage{natbib}
\usepackage{multirow}  
\usepackage{lineno} 

\newcommand{\be}{\begin{eqnarray}}
\newcommand{\ee}{\end{eqnarray}}

\newcommand{\ba}{\begin{align}}
\newcommand{\ea}{\end{align}}

\makeatletter

\newcommand{\Rmnum}[1]{\expandafter\@slowromancap\romannumeral #1@}
\makeatother

\begin{document}
\title{Critical dynamics and its interferometry in the one-dimensional p-wave-paired Aubry-André-Harper model}

\author{Zhi-Han Zhang}
\author{Han-Chuan Kou}

\author{Peng Li}
\email{lipeng@scu.edu.cn}

\affiliation{College of Physics, Sichuan University, 610064, Chengdu, People’s Republic of China}



\begin{abstract}

    In this work, we focus on the critical dynamics of the one-dimensional quasiperiodic p-wave-paired Aubry-André-Harper model, which exhibits a transition point between the gapped critical and the gapless localized phases. First, we disclose that the dynamical exponent ‌features two distinct plateaus in the gapless localized phase. Besides the plateau with dynamical exponent $z=1.388$ near the transition point, there is a second one with $z=1$ away from the transition point. Then we demonstrate that these two plateaus intrinsically affect the critical dynamics with moderate quench rate by employing both one-way and round-trip quench protocols. In the one-way quench protocol, we clarify that the Kibble-Zurek (KZ) regime consists of two sub-regimes with different KZ exponents, which is a direct consequence of the two-plateau structure of the dynamical exponent. We also diagnose the KZ, presaturated, and saturated regimes by varying the quench rate from slow to fast limits. While, in the round-trip quench protocol, we confirm the interference effect of two critical dynamics and show a narrow neck in the oscillatory density of defects. It turns out that the position of the neck depends on the ratio of to-and-fro quench rates and the p-wave superconductivity pairing amplitude. We show how the position of the neck can be used to quantitatively determine the turning point of the two KZ sub-regimes.

\end{abstract}

\maketitle


\section{Introduction}\label{Section:Intro}
    The renowned Kibble-Zurek mechanism (KZM) was first proposed to understand the signatures of phase transitions that have occurred in the early universe, determining the density of defects left in the broken symmetry phase \cite{Kibble1976, Kibble1980}. Soon later, similar critical dynamics in condensed matter systems is uncovered to be described by KZM also \cite{Zurek1985}. Specifically, for a linear quench protocol, the distance from the critical point can be parameterized by $\epsilon(t)=(t-t_c)/\tau_Q$, where $\tau_Q$ is the quench time and $t_c$ marks the time when the critical point is crossed. After crossing the critical point, the final density of defects exhibits the KZ scaling law, $\sim\tau_Q^{-dv/1+vz}$, where $d$ is the dimension of space, $\nu$ and $z$ are critical exponents. Since then, progresses have been achieved in both theoretical and experimental researches \cite{Dziarmaga2005, Fischer2006, Fischer2007, Fischer2008, Fischer2010, Campol2016, Keesling2019, Campol2022, Bacsi2023, Hetenyi2024}. A recent study revealed an interference effect of two successive critical dynamics realized by a properly designed quench protocol, which can be easily observed through the oscillatory KZ behavior in the defect density \cite{Kou2022}. For fast quench rates, the KZ scaling law will be broken. However, extensive studies have revealed that there are saturated and presaturated regimes beyond the KZ regime \cite{Campo2010, Griffin2012, Sonner2015, Chesler2015, Campo2019, Sun2021, Campo2023PRL, Campo2023PRB, Campo2023PRD, Kou2023}.

    On the other hand, the quantum localization was initially found in a \emph{randomly disordered} system \cite{Anderson1958}. Since then, the research on the topic of localized-delocalized phase transition has attracted a lot of attention \cite{Evers2008, Martin2006, Sarma2015, Semeghini2015, Delande2017, Hetenyi2021, Wang2023, Chen2024}. Later, a similar localized-delocalized phase transition was found to be realized in a \emph{quasiperiodic disordered} system, the so-called Aubry-André-Harper (AAH) model for one-dimensional wires \cite{Aubry1980, Harper1955}. It has stimulated a lot of theoretical and experimental researches on quasiperiodic systems, in which intriguing phenomena such as fractal spectrum, many-body localization, and mobility edge are revealed \cite{Luck1986, Doria1988, Doria1989, Benza1989, Satija1989, Luck1993, Joachim1997, Hermisson1998, Hermisson2000, Negro2003, Fallani2007, Roati2008, Lahini2009, Modugno2009, Kraus2012, Segev2013, Sanchez2014, Sanchez2019, Roy2021, Roosz2020, Sanchez2022, Yu2024}. The critical dynamics of the AAH model have also been extensively studied, which shows a significant difference from that of the randomly disordered systems \cite{Roosz2014, Dziarmaga2019, Roosz2024}.

    Recently, the descendant p-wave-paired AAH model attracts much attention \cite{Cai2013, Wang2016, Chen2016, Chandran2018, Gao2021, Gao2023, You2022-PRA, You2022-PRB, Roy2024}. A variety of transitions between delocalized, critical, and localized phases had been revealed in this quasiperiodic system \cite{Cai2013, Wang2016}. Then the quantum criticality was disclosed to be intermediate to that of the clean and randomly disordered systems \cite{Chandran2018}. Later, the unconventional critical exponents have been unraveled through the critical behavior of localized-critical and critical-delocalized phase transitions \cite{Gao2021, You2022-PRA, You2022-PRB}. Although the slow and sudden critical dynamics have been investigated, the quench dynamics within the regime of fast quench rate still lacks thorough research.

    In this work, we focus on the AAH model with p-wave SC paring. First, we extract the dynamical exponent $z(\epsilon)$ in the gapless localized phase by the scaling behavior of the energy gap, $\Delta_g \sim L^{-z(\epsilon)}$, where $L$ is the system's size and $\epsilon=V - V_c$ measures the distance from the critical point. We observe two plateaus‌ in the $z(\epsilon)$ profile: one near the critical point ($z\approx1.388$ at $\epsilon\approx0$), and the other away from the critical point ($z\approx 1$ at $\epsilon\gtrsim1$). Then we demonstrate that the critical dynamics of the system are modulated by the two plateaus according to the adiabatic-impulse-adiabatic approximation. The plateau with $z\approx1.388$ governs the critical dynamics with slow enough quench rate since it is near the transition point. While the other with $z\approx1$ determines the behavior of the critical dynamics with moderately fast quench rate.

    We employ two quench protocols, say, the one-way and round-trip linear ones, to uncover the influence of the dynamical exponent. In the one-way quench protocol, we diagnose the KZ, presaturated, and saturated regimes. Notably, we find that the KZ regime can be divided into two sub-regimes, the so-called KZ-I and KZ-II, with different KZ exponent. In the round-trip quench protocol, we observe a narrow neck in the oscillatory density of defect as the interference effect of two successive critical dynamics \cite{Kou2022} and show the position of the neck can be utilized to distinguish the KZ-I and KZ-II regimes.

    The rest of the paper is organized as follows: In \ref{Section:Model-Hamiltonian}, we briefly introduce the model Hamiltonian. In \ref{Section:dynamical-Z}, we demonstrate the two-plateau structure of the dynamical exponent. In \ref{Section:quenchdynamics}, we employ two different quench dynamics, the one-way and round-trip quench protocol, to unravel the impact of the dynamical exponent on the critical dynamics. At last, we give a summary.

 \begin{figure}[t]
      \begin{center}
    		\includegraphics[width=3.2in,angle=0]{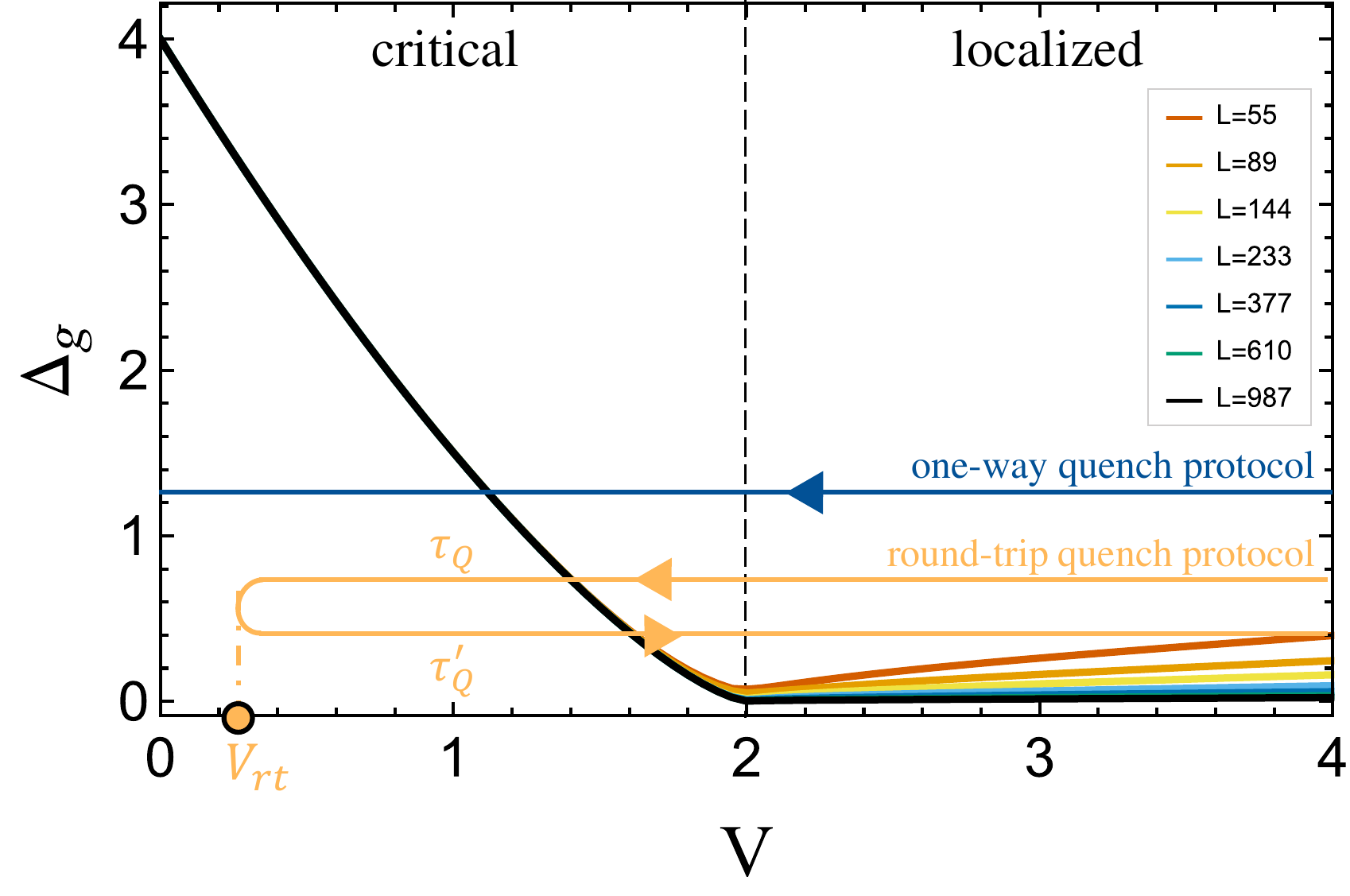}
      \end{center}
      \caption{The energy gap behavior and the quench protocols at the p-wave paring amplitude $\Delta = 1$. The energy gap $\Delta_g$ is collected as a function of potential strength $V$ by varying the size of system $L$. In the localized phase, the gap gradually closes with the increasing $L$. Moreover, two quench protocols are marked by different arrows. The one-way quench protocol linearly drives the system from the localized phase to the critical phase. The round-trip one consists of two successive linear ramps with the turning point $V_{rt}$, $\tau_Q$ and $\tau_{Q}^{\prime}$ denote quench times of two linear ramps respectively.}
      \label{fig:spectrum}
  \end{figure}

\section{Model Hamiltonian}\label{Section:Model-Hamiltonian}

    The AAH model with p-wave superconductive pairing is described by the non-interaction Hamiltonian:
    \begin{equation}
        H = \sum\limits_{j=1}^{L} -J c_{j+1}^{\dagger}c_{j}+ \Delta c_{j+1}^{\dagger}c_{j}^{\dagger} + h.c. + V_j c_{j}^{\dagger} c_{j} ,\label{eq:halAASC}
    \end{equation}
    where $c_j^{\dagger}\left(c_j\right)$ is creation(annihilation) operator of fermion. Here, $J$ is the nearest-neighbor hopping amplitude, and $\Delta$ is the p-wave pairing amplitude. The incommensurate potential
    \begin{equation}
        V_j = 2 V\cos(2\pi \gamma j + \phi),
    \end{equation}
    varies at each lattice site with a random phase $\phi \in [ 0,2\pi )$ is introduced as a mean to average over the quasiperiodic potential field. The irrational number $\gamma = \left(\sqrt{5}-1\right)/2$ is the limit of the ratio of two Fibonacci number $F_n/F_{n+1}$, which ensures system satisfying the periodic boundary condition, i.e. $c_{j+L}=c_j$.

    The Hamiltonian can be rewritten as a $2L \times 2L$ matrix,
    \begin{equation}
    \small{
    \begin{pmatrix}
      V_1 & -J & 0 & \cdots & -J & 0 & -\Delta & 0 & \cdots & \Delta \\
      -J & V_2 & -J & \ddots & \vdots & \Delta & 0 & -\Delta & \cdots & 0 \\
      0 & -J & \ddots & \ddots & 0 & 0 & \Delta & \ddots & \ddots & \vdots \\
      \vdots & \ddots & \ddots & V_{L-1} & -J & \vdots & \vdots & \ddots & 0 & -\Delta \\
      -J & \cdots & 0 & -J & V_L & -\Delta & 0 & \cdots & \Delta & 0 \\
      0 & \Delta & 0 & \cdots & -\Delta &  -V_1 & J & 0 & \cdots & J  \\
      -\Delta & 0 & \Delta & \cdots & 0 & J & -V_2 & J & \ddots & \vdots \\
      0 & -\Delta & \ddots & \ddots & \vdots & 0 & J & \ddots & \ddots & 0 \\
      \vdots & \vdots & \ddots & 0 & \Delta & \vdots & \ddots & \ddots & -V_{L-1} & J \\
      \Delta & 0 & \cdots & -\Delta & 0 & -J & \cdots & 0 & -J & V_L
    \end{pmatrix}},
    \end{equation}
    in the bases
    \[
        \left(c_1,\cdots,c_L,c_1^\dagger,\cdots,c_L^\dagger\right)^T,
    \]
    According to the Bogoliubov-de Gennes(BdG) transformation,
    \begin{equation}
        \eta_j = \sum\limits_{m=1}^{L}u^{*}_{m,j}c_{m}+v^{*}_{m,j}c_{m}^\dagger, \label{eq:insBdGtrans}
    \end{equation}
    Hamiltonian is diagonal:
    \begin{equation}
        H = \sum\limits_{j=1}^{L} 2 \omega_j \eta_j^\dagger \eta_j-\sum\limits_{j=1}^{L}\omega_j, \label{eq:diagHal}
    \end{equation}
    where $(u_{i,j},v_{i,j})$ is the BdG modes transformation matrix and $0 < \omega_j < \omega_{j+1}$ are energy of single quasiparticle states. The qusiparticles $\eta_j$ together constitute a vacuum annihilated by any $\eta_j$, which is the ground state of this system.

    The phase diagram of the AAH model with p-wave superconducting pairing amplitude $\Delta$ and the quasi-periodic potential strength $V$ has been thoroughly investigated in previous research \cite{Cai2013, Wang2016, Gao2021}. Three phases, named extended, critical, and localized, are explored. The system undergoes the extended-critical phase transition at $V=\left| J-\Delta \right|$ and the critical-localized phase transition at $V=\left| J+\Delta \right|$.

\begin{figure}[t]
    \begin{center}
    	\includegraphics[width=3.2in,angle=0]{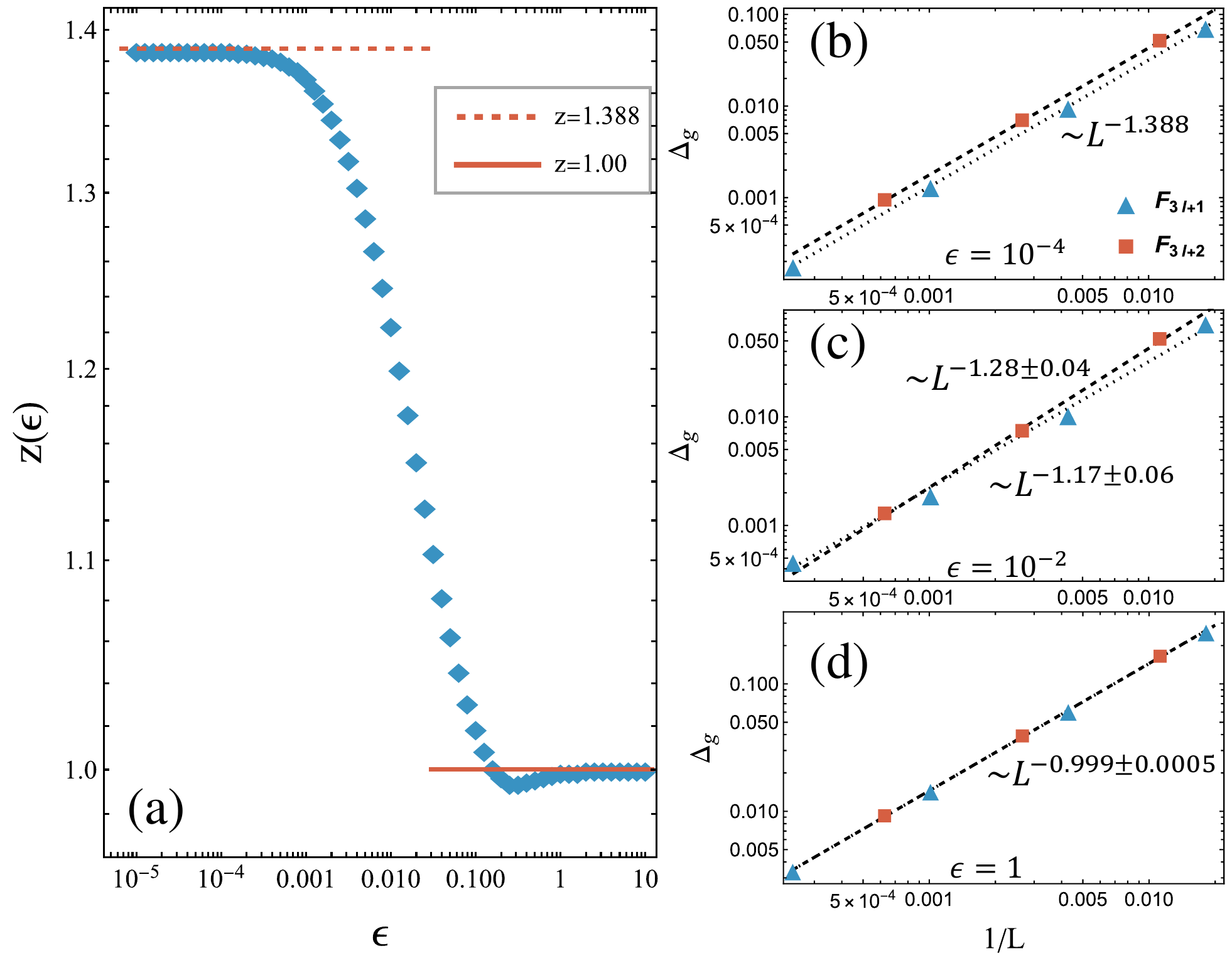}
    \end{center}
    \caption{The dynamical exponent in the gapless localized phase. (a) The dynamical exponent $z$ as a function of the distance from the critical point $\epsilon$. Two plateaus at $z\approx1$, and $1.388$, emerge at far and near from the critical point ($\epsilon=0$). The fitted straight lines for each subsequences gives: , (b) $\Delta_g \sim L^{1.3872\pm0.0003}$ and $ L^{1.3877\pm0.0005}$ with $\epsilon=10^{-4}$, (c) $\Delta_g \sim L^{1.28\pm0.04}$ and $ L^{1.17\pm0.06}$ with $\epsilon=10^{-2}$, and (d) we obtained $\Delta_g \sim L^{0.999\pm0.0005}$ with $\epsilon=1$. These conclusions were derived from analyzing 500 different values of $\phi$.}
    \label{fig:dynamicalexponent}
\end{figure}

\section{Dynamical exponents}\label{Section:dynamical-Z}
    In this section, we analyze the critical behavior of the critical-localized phase transition and unravel that the dynamical exponent is continuously dependent on the distance $\epsilon$ in the gapless localized phase.

    The critical exponent, including the dynamical exponent $z$ and correlation-length exponent $\nu$, characterize the universal scaling behavior near a phase transition. The correlation length exponent for the critical-localized phase transition been obtained in past research \cite{Gao2021, You2022-PRA} with exact value $\nu=1$, which shows that the correlation length $\xi$ diverges with a scaling behavior $\xi \sim \epsilon^{-1}$, where $\epsilon = V-V_c$ is the distance from the critical point.

    However, the dynamical exponent $z$ is found to be a continuous function of the quasiperiodic potential strength in the gapless localized phase.

\subsection{Dynamical exponent at $\Delta = 1$}
     We set the p-wave pairing amplitude $\Delta = 1$ and the hopping amplitude $J = 1$ to investigate the dynamical exponent for the critical-localized phase transition. In this case, the critical point of this phase transition is located at $V_c= \left|J+\Delta\right| = 2$. Interestingly, the localized phase is found to be gapless within a large potential strength $V>V_c$ (Fig. \ref{fig:spectrum}), which is characterized as a gap
    \begin{equation}
        \Delta_g \sim L^{-z}, \label{eq:gap-L}
    \end{equation}
    with $\Delta_g = 2(\omega_1+\omega_2)$ being twice the sum of the two lowest positive energies \cite{Caneva2007, Gao2021}.

    Figure \ref{fig:dynamicalexponent}(a) shows how the dynamical exponent $z$ depends on the distance from the critical point $\epsilon$ in the gapless localized phase. The exponent $z$ is determined by averaging the linear fits of the scaling relation $\Delta_g \sim L^{-z}$ across three Fibonacci subsequences: $F_{3l}$, $F_{3l+1}$, and $F_{3l+2}$, following the methodology established in Refs. \cite{Ino2006, You2022-PRB}. In Fig. \ref{fig:dynamicalexponent}(a), the plateau $z_\text{I} \approx 1$ holds with large $\epsilon$ and then gradually rises, eventually saturating at $z_\text{II} \approx 1.388$ as $\epsilon$ narrow, aligning with the dynamical exponent obtained at the critical point \cite{Gao2021, You2022-PRA}. The dependence of the dynamical exponent on the potential strength in the gapless localized phase is similar to that of the random disorder system \cite{Dziarmaga2006, Caneva2007}. However, unlike the random disorder system, $z$ does not diverge at the critical point but instead saturates to a constant. The gap as a function of the size of the system $L$ by varying the distance from the critical point is shown in Fig. \ref{fig:dynamicalexponent}(b), (c), and (d). In Fig. \ref{fig:dynamicalexponent}(b), near the critical point, dynamical exponents $z = 1.3872\pm0.0003$ and $1.3875\pm0.0005$ are obtained from different subsequences with a negligible difference. However, this difference becomes inescapable, $z = 1.28\pm0.04$ and $1.17\pm0.06$, with the distance of the critical point gradually increasing, see Fig. \ref{fig:dynamicalexponent}(c). When the potential strength is further away from the phase transition point, different subsequences will share a common power-law exponent $z=1$ (Fig. \ref{fig:dynamicalexponent}(d)).

\subsection{Dynamical exponent for varying $\Delta$}
    The dynamical exponent in the localized phase is also investigated by varying $\Delta$ in Eq. (\ref{eq:halAASC}) with the hopping strength $J=1$.

\begin{figure}[t]
    \begin{center}
    	\includegraphics[width=3.0in,angle=0]{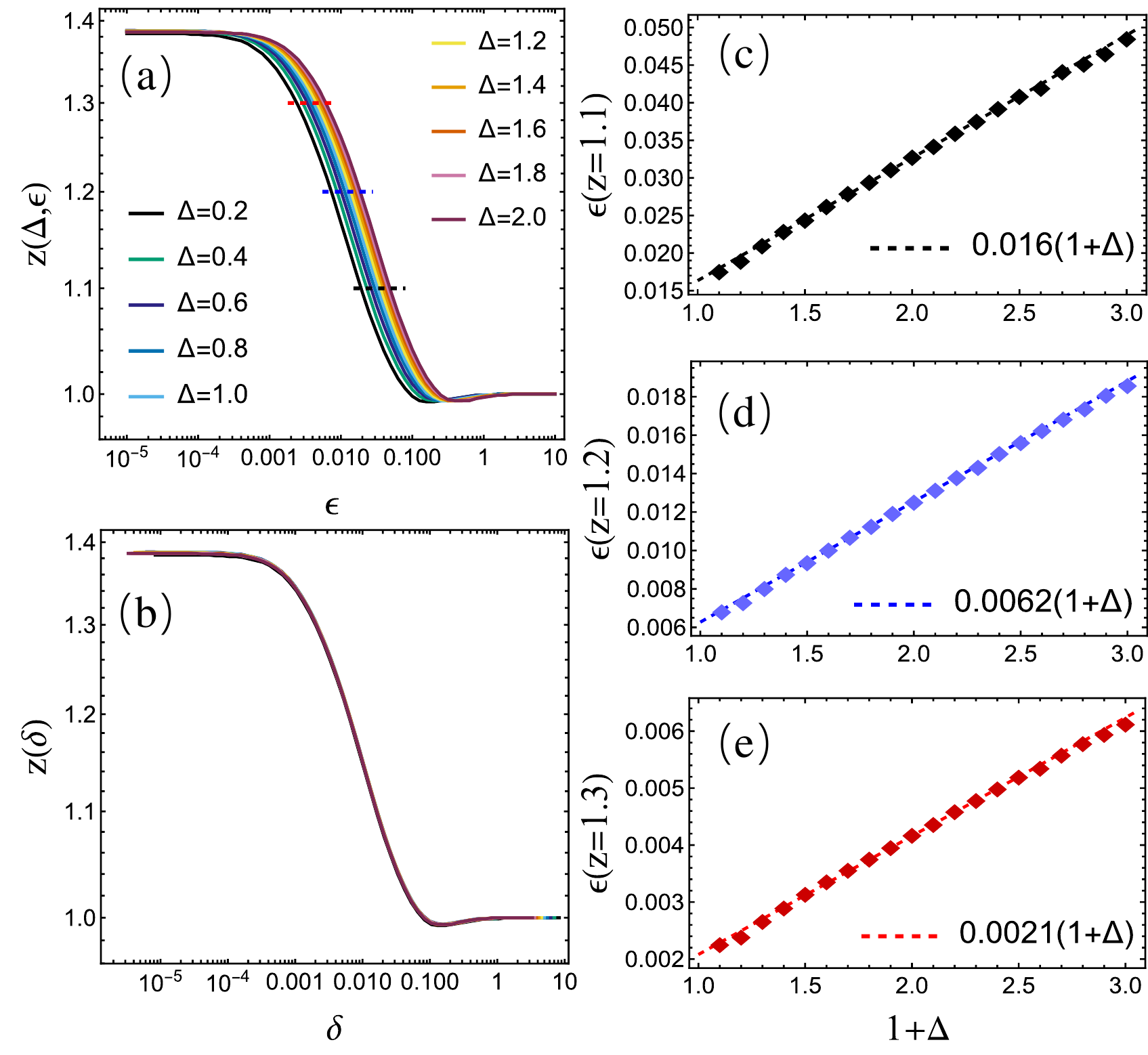}
    \end{center}
    \caption{(a) Dynamical exponent as a function of distance from the critical point, $\epsilon=V-V_c$. (b) Dynamical exponent as a function of reduced potential strength $\delta = (V-V_c)/V_c$, with $V_c=1+\Delta$. All curves collapse into a universal scaling with two distinct plateaus. In (c), (d), and (e), we show linear relationships of the distance $\epsilon$ and $V_c$ for a certain value of the dynamical exponent. (c) $\epsilon = 0.016V_c$ with $z=1.1$. (d) $\epsilon = 0.0063V_c$ with $z=1.2$. (e) $\epsilon = 0.0021V_c$ with $z=1.3$. These conclusions were derived from analyzing 500 different values of $\phi$.}
    \label{fig:DE02}
\end{figure}

    In Fig. \ref{fig:DE02}(a), the dynamical exponent $z(\Delta, \epsilon)$ in the localized phase is collected as a function of $\epsilon$, with the critical point of the critical-localized phase transition $V_c = |J+\Delta|=|1+\Delta|$. Figure \ref{fig:DE02}(c), (d), and (e) show how the distance from the critical point $\epsilon$ in Fig. \ref{fig:DE02}(a) depends on the p-wave pairing amplitude. Here, $\epsilon(z,\Delta)$ is obtained as the distance from the critical point at which the dynamical exponent falls below a threshold, $z(\Delta,\epsilon)=z_0$. According to the numerical result by setting $z_0=1.1$, $1.2$ and $1.3$, we show that $\epsilon(z,\Delta)$ is a direct proportional function of $V_c$:
    \begin{equation}
        \epsilon(z_0,\Delta) = \alpha(z_0)V_c,
    \end{equation}
    where $\alpha(z_0)$ is a proportional parameter related to $z_0$. This relation discloses that the dynamical exponent $z$ should be a function of the reduced potential strength $\delta = (V-V_c)/V_c$:
    \begin{equation}
        z(\Delta, \epsilon) = z(\delta). \label{eq:z-delta}
    \end{equation}
    The dynamical exponent as a function of the reduced potential strength is shown in Fig. \ref{fig:DE02}(b), where the curves with different $\Delta$ collapse into a universal scaling. Two plateaus are observed in the $z(\delta)$ profile, one near and the other far from the critical point.

    In conclusion, the dynamical exponent is continuously dependent on the reduced potential strength in the gapless localized region, with two plateaus at $z=1$ and $1.388$:
    \begin{equation}
    z(\delta) = \left\{
    \begin{aligned}
    &1, &(\delta \gtrsim 1) , \\
    &1.388, &(\delta \approx 0). \label{eq:dynamicalexponent}
    \end{aligned}
    \right.
    \end{equation}

\section{Quench Dynamics}\label{Section:quenchdynamics}
    The quench dynamics are used to investigate the critical behavior of a second-order phase transition by properly customizing the quench protocol. In this section, we use two different quench protocols, the one-way and round-trip one, to disclose the behavior of the dynamical exponent in the gapless localized phase.

\subsection{One-way quench protocol}
    The one-way quench protocol is chosen to be inquired first. In this protocol, the distance from the critical point can be parameterized as follows:
        \begin{equation}
            \epsilon(t) = -\frac{t}{\tau_Q},\label{eq:linearquench}
        \end{equation}
    where $\tau_Q$ is the quench time.

   During the slow quench process with a large $\tau_Q$, the adiabatic-impulse-adiabatic approximation begins to take effect. Initially, the system is prepared in a localized ground state with $t_i \rightarrow -\infty$. The system undergoes adiabatic evolution when the potential strength is far away from the critical point. Then, adiabaticity breaks down when the inverse transition rate $\left|\epsilon/\dot{\epsilon}\right|$ equals the relaxation time $\tau \sim \Delta_g^{-1} \sim |t/\tau_Q|^{-z \nu}$ at $t = -\hat{t}$. In this case, $\hat{t}$ scales as:
        \begin{equation}
        \hat{t} \sim \tau_Q^{\frac{\nu z(\hat{\epsilon})}{1+\nu z(\hat{\epsilon})}},\label{eq:hattime}
        \end{equation}
        corresponding to
        \begin{equation}
        \hat{\epsilon} \sim \tau_Q^{-\frac{1}{1+\nu z(\hat{\epsilon})}}, \label{eq:hateps}
        \end{equation}
    where the dynamical exponent $z(\hat{\epsilon})$ is obtained from Eq. (\ref{eq:gap-L}). At $\epsilon = -\hat{\epsilon}$, the ground state gets excited and freeze-out with a corresponding correlation length $\hat{\xi} \sim \hat{\epsilon}^{-\nu}$. The state does not change until $t = \hat{t}$ when the relaxation time keeps pace with the evolving parameter successfully. After $\hat{t}$, the system gets adiabatic evolution, and the correlation length corresponding to $-\hat{t}$ is expected to populate. Consequently, the density of excitation probability for a one-dimensional system scales as follows:
\begin{equation}
            n_\text{ex} \simeq \hat{\xi}^{-1} \sim \tau_Q^{-\frac{\nu}{1+\nu z(\hat{\epsilon})}}, \label{eq:KZpowerlaw}
\end{equation}
    which predicts the scaling law of defect density depends on the quench time. Moreover, the KZ scaling law suggests that the correlation length $\hat{\xi}$ and freeze-out time $\hat{t}$ scale diverge in the adiabatic limit and become the only scales in the long-wavelength limit \cite{Dziarmaga2010, Zurek2016}. Deducing by this logic, the hypothesis shows that in the impulse region, the evolution of the density of excitation can be expressed as:
         $n_\text{ex}(t) \sim \hat{n}_{ex} F\left(t/\hat{t}\right)$ or
        \begin{equation}
            n_\text{ex}(t) \cdot \tau_Q^{\frac{\nu}{1+\nu z(\hat{\epsilon})}} \sim F\left(t/\hat{t}\right), \label{eq:time-dependentFunction}
        \end{equation}
    where $F$ is a non-universal scaling function \cite{Deng2008,David2012,Zurek2022}. Consequently, the scaled excitation density has a scaling function dependent on the scaled time $t/\hat{t}$.

    However, as the quench speed gradually increases, the KZ scaling law gradually fails because of the adiabatic-impulse-adiabatic approximation failure. In the fast quench, the defect behavior can be divided into saturated and pre-saturated regimes. The saturated regime (SR) is defined by an extremely fast quench rate, where the KZ scaling law breaks down. In this regime, the defect density forms a plateau with $n_\text{ex}=n_\text{su}$. This plateau is attributed to sudden quench and depends on the initial potential strength $V_i$ \cite{Sun2021}. For the transverse Ising model, the defect density within the SR scales as a function of $\tau_Q^{2}$ :
    \begin{equation}
     \begin{split}
         n_\text{ex} &= n_{su } - K V_{i}^{3} \tau_{Q}^{2}, \\
         n_\text{su} &= \frac{1}{2} - \frac{\alpha}{V_i},\label{eq:saturate-d}
     \end{split}
    \end{equation}
where $K$ and $\alpha$ are fitting parameters. When the quench rate declines and reaches the tuning point, $\tau_Q=\tau_Q^{S}$, the pre-saturate regime (PSR) appears because the initial potential strength is out of the impulse regime \cite{Kou2023}. The defect's density within the PSR is a linear combination of $\tau_Q^{1/2}$ and $\tau_Q^{3/2}$:
    \begin{equation}
         n_\text{ex} = \frac{1}{2} - A\tau_{Q}^{1/2} + B\tau_{Q}^{3/2},\label{eq:presaturate-d}
    \end{equation}
where $A$ and $B$ are fitting parameter related to $V_i$.

To obtain the final defects, it is helpful to introduce the time-dependent BdG transformation,
    \begin{equation}
    \gamma_j = \sum\limits_{i=1}^{L}u^{*}_{j,i}(t)\tilde{c}_{i}+v^{*}_{j,i}(t)\tilde{c}_{i}^\dagger,\label{eq:time-dependentBdG}
    \end{equation}
    where $\gamma_j$ is the quasi-particle that can annihilate the initial ground state. The time-dependent Bogoliubov modes $(u_{i,j}(t),v_{i,j}(t))$ follow the Heisenberg equation $i\hbar \frac{\partial}{\partial t} \tilde{c}_j= \left[\tilde{c}_j, U^{\dagger}H U\right]$ with $\tilde{c_j}=U^{\dagger}c_j U$, and $U$ is the time-evolve operator. Notably, $\gamma_j$ is independent of time, $\frac{d}{dt}\gamma_j=0$, because it is defined by the initial ground state. According to the Heisenberg equation and property of $\gamma_j$, the dynamical version of the BdG equation can be written as,
    \begin{equation}
    \small{ \left\{\begin{split}
    i \frac{\partial}{\partial t} u_{ji}(t) &= V_j(t) u_{ji}(t)- J(u_{j,i-1}(t) + u_{j,i+1}(t))\\
    &+\Delta(v_{j,i-1}(t) - v_{j,i+1}(t)), \\
    i \frac{\partial}{\partial t} v_{ji}(t) &= -V_j(t) v_{ji}(t) + J(v_{j,i-1}(t) + v_{j,i+1}(t)) \\
    &- \Delta(u_{j,i-1}(t) - u_{j,i+1}(t)),  \label{eq:time-dependentBdG}
    \end{split} \right. }
    \end{equation}
    with initial conditions $u_{j,i}(t_0) = u_{j,i}$, $v_{j,i}(t_0) = v_{j,i}$ where $(u_{j,i}, v_{j,i})$ is the instantaneous BdG modes defined as (\ref{eq:insBdGtrans}) at $V_j = V_j(t_0)$ in Schrodinger picture. In this case, the density of excited probability can be obtained as:
    \begin{equation}
    \small{
    \begin{split}
    n_\text{ex}(t) &= \frac{1}{L}\sum\limits_{j=1}^{L} p_{j}^{ex} \\
    &= \frac{1}{L}\sum\limits_{j=1}^{L}\langle \psi(t) |\eta_j^{\dagger} \eta_j |\psi(t) \rangle \\
    &= \frac{1}{L}\sum\limits_{j=1}^{L}\sum\limits_{n=1}^{L} \beta_{j,n}\beta^{*}_{j,n} \langle \psi(t_0) |\gamma_n \gamma^{\dagger}_n |\psi(t_0) \rangle \\
    &= \frac{1}{L}\sum\limits_{j=1}^{L}\sum\limits_{n=1}^{L}\left| \sum\limits_{k=1}^{L} v_{j,k}u_{n,k}(t) + u_{j,k}v_{n,k}(t) \right|^2,\label{eq:defects}
    \end{split}}
    \end{equation}
    with
    \begin{equation}
    U^{\dagger}\eta_m U =\sum\limits_{n=1}^{L} \alpha_{m,n} \gamma_{n} + \beta_{m,n} \gamma_{n}^{\dagger},
    \end{equation}
    where $|\psi(t) \rangle$=$U(t,t_0)|\psi(t_0) \rangle$, and $|\psi(t_0) \rangle$ represents a Bogoliubov vacuum corresponding to the initial ground state, annihilate by the quasi-particle annihilate operator $\gamma_m$ defended by time-dependent BdG transformation $(\ref{eq:time-dependentBdG})$.

\begin{figure}[t]
      \begin{center}
    		\includegraphics[width=3.2in,angle=0]{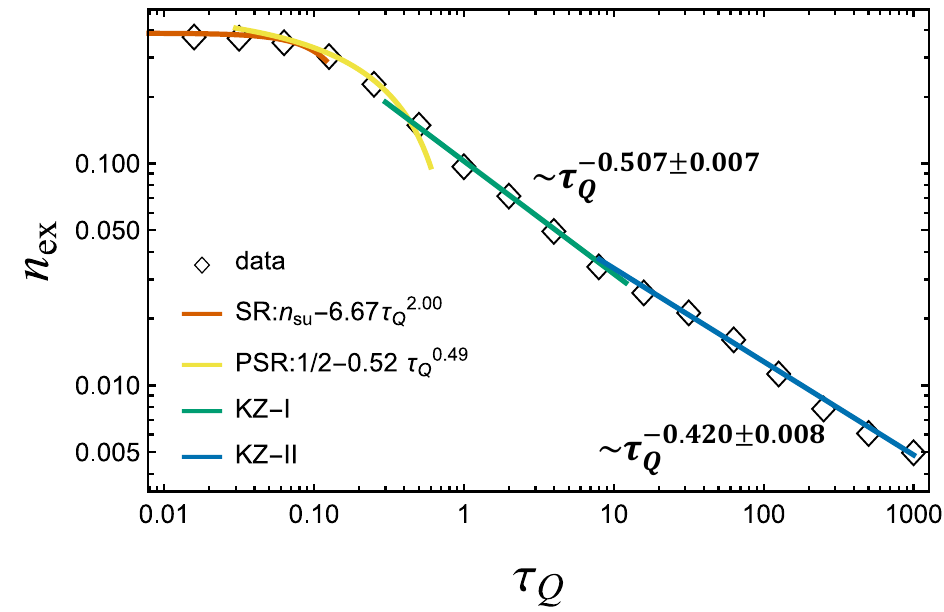}
      \end{center}
      \caption{The defects as a function of quench time $\tau_Q$ with system size $L=987$. The final defects are obtained from the linear quench by adjusting the quench rate from slow to very fast. After the fast quench regimes, the fit gives $n_\text{ex} \sim \tau_Q^{-0.507\pm0.007}$ (KZ-I) and $n_\text{ex} \sim \tau_Q^{-0.420\pm0.008}$ (KZ-II), corresponding to the fast and slow quench rate. The alternation of the KZ exponents reflects the two plateaus of the dynamical exponent $z \approx 1$ and $z \approx 1.388$. At saturate regime (SR), the density of defects is saturated to a plateau at first and then decreases with a scaling $n_\text{ex}\sim \tau_{Q}^{2}$. In the pre-saturate regime (PSR), $n_\text{ex}$ declines as $1/2-0.52\tau_Q^{0.49}$ until $\tau_Q > 1$.}
      \label{fig:owquench}
\end{figure}

 \begin{figure}[t]
      \begin{center}
    		\includegraphics[width=3.2in,angle=0]{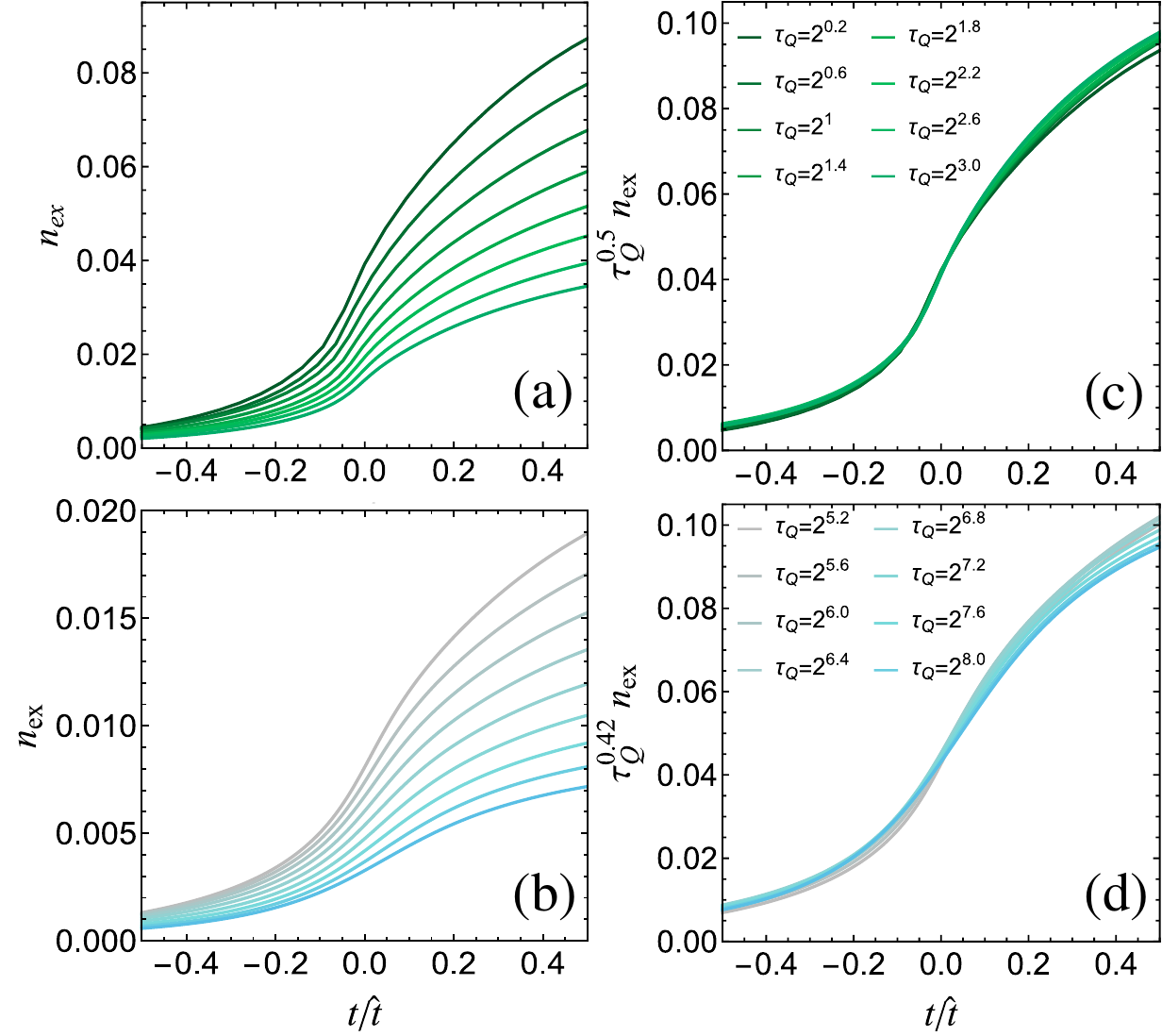}
      \end{center}
      \caption{The non-universal function in the impulse regime with fast and slow quench rate: the density of defect $n_\text{ex}$ as a function of the scaled time $t/\hat{t}$ with (a) the quench rate belonging in KZ-I and (b) the quench time belonging in KZ-II. According to multiplying $n_\text{ex}$ by KZ scaling law of KZ-I and KZ-II, curves with different quench rates in (c) and (d) collapse into a universal curve, respectively.}
      \label{fig:defectsfunction}
 \end{figure}

\subsubsection{Final defects after the one-way quench}
     The linear quench described by Eq. (\ref{eq:linearquench}) is utilized to analyze the behavior of the final defect's density populated after the system quenching through the critical-localized phase transition point with the quench time $\tau_Q$. By setting the system parameter $\Delta, J = 1$, the corresponding one-way quench protocol is shown in Fig. \ref{fig:spectrum}. We consider a deeply localized initial state at $t_i = -V_i\tau_Q$ and drive the system across the critical point at $t = 0$. Then, the system stop at a final time $t_{f} = 2\tau_Q$, corresponding to $V = 0$, with a defects' density $n_\text{ex}(t_{f})$.

    The final defects density $n_\text{ex}$ as a function of the quench time $\tau_Q$ is shown in Fig. \ref{fig:owquench} from slow to very fast quench rate with the initial potential strength $V_i = 4$.  According to the behavior of the defect density, we indicate the saturated, pre-saturated, and Kibble-Zurek regimes. In the saturated regime (SR), the saturate plateau $n_\text{su}=0.3865$ appears when the quench rate is very fast, as the dashed red line. A fitting of $(n_\text{su}-n_\text{ex})\sim\tau_Q^{s}$ yields $s=2.004\pm0.005$, which is in good agreement with Eq. (\ref{eq:saturate-d}). With the increasing of quench time, the initial potential strength $V_i$ is out of the non-adiabatic regime, $V_i>V+\hat{\epsilon}$, and quench dynamics turns into pre-saturated regime (PSR). In this case, $n_\text{ex}$ diminishes as the expression $1/2 - 0.52\tau_Q^{0.49}$, which is revealed by a fitting of $(1/2-n_\text{ex})=\tau_Q^{s}$ yields $s=0.490\pm0.009$. The $\tau_{Q}^{3/2}$ term in Eq. (\ref{eq:presaturate-d}) does not appear because of the small fitting parameter $B$. After the quench time gets to $\tau_Q \approx 1$, the adiabatic-impulse-adiabatic hypothesis kicks in, and the KZ scaling law can reasonably predict the behavior of defects' density.

    In the KZ regime, the power-law fit and Eq. (\ref{eq:KZpowerlaw}) imply two dynamical exponents $z_{\text{I}}^{D}\approx0.972\pm0.027$ and $z_{\text{II}}^{D}\approx1.381\pm0.045$ for exact $\nu=1$, corresponding to the moderately fast and slow quench rates. The two values of dynamical exponents match the ones obtained by the finite-size scaling as shown in Fig. (\ref{fig:dynamicalexponent}), $z_\text{I}=1$ and $z_\text{II}=1.388$, with systematic errors $3\%$ and $0.6\%$ respectively. In the moderately fast quench rate, $z_\text{I}$ closely resembles that of the transverse field Ising model. In the slow quench regime, the KZ exponent shifts to $z_\text{II}$, which is in good agreement with the known result \cite{Gao2021}. ‌Specifically, for the moderately fast quench rate, the impulse regime is far from the critical point but does not extend beyond $V_i$ with the dynamical exponent staying at a constant $z(\hat{\epsilon}\gtrsim1)\approx1$. In this case, the defect's density behaves as a power-law scaling $n_\text{ex}\sim\tau_Q^{-0.5}$. With the quench rate slowing down, the non-adiabatic regime becomes narrow and near the critical point. The dynamical exponent will saturate to a plateau $z(\hat{\epsilon}\rightarrow 0)\approx1.388$, corresponding to the excitation probability scaling as $n_\text{ex}\sim\tau_Q^{-0.42}$, which in good agreement with the result in the past research \cite{You2022-PRA}. This variable KZ exponent is similar to that in the random Ising chain affected by the Griffith phase \cite{Dziarmaga2006, Caneva2007}. However, the KZ exponent will be steady because the dynamical exponent saturates to a constant at the critical point instead of diverging.

    The influence of the dynamical exponent is also reflected in the non-universal function (\ref{eq:time-dependentFunction}), which discloses a universal behavior of the excitation probability in the impulse region. The defect's density as a function of scaled time in the non-adiabatic region is shown in Fig. \ref{fig:defectsfunction}(a) and (c) by varying quench time $\tau_Q$ belonging to KZ-I and KZ-II, respectively. Different scaling factors, $\tau_Q^{0.5}$ and $\tau_Q^{0.42}$ corresponding to the inverse of $\hat{n}_\text{ex}$, are multiplied in $n_\text{ex}$ to disclosed the non-universal function (\ref{eq:time-dependentFunction}) in Fig. \ref{fig:defectsfunction}(c) and (d).

\subsubsection{Final defects with fast quench rate}
\begin{figure}[t]
      \begin{center}
    		\includegraphics[width=3.2in,angle=0]{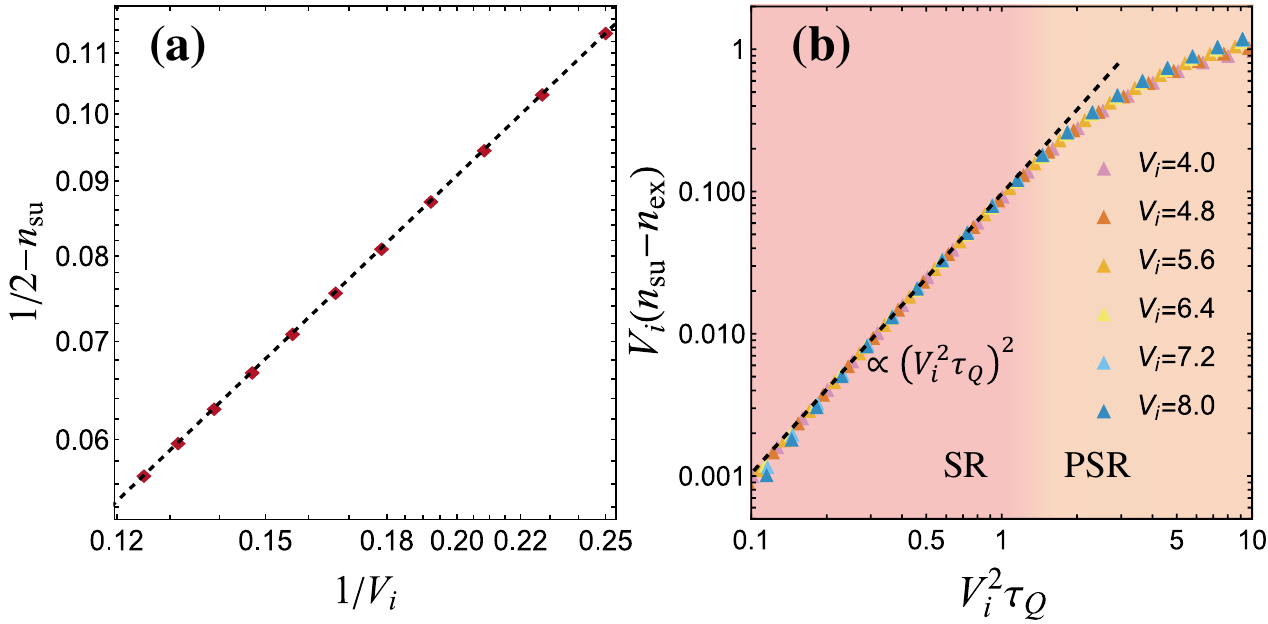}
      \end{center}
      \caption{The scaling behavior in the fast quench regime. (a) the reduced saturate plateau $1/2-n_\text{su}$ versus the reciprocal of initial potential strength $1/V_{i}$. A linear model fitting for $1/2-n_\text{su}=\alpha V_i$ yields the fitting parameter $\alpha = 0.425\pm0.0004$. (b) $(n_\text{su}-n_\text{ex})V_i$ versus $V_i^2 \tau_Q$ with varying $V_i$. The numerical data collapses to the power-law scaling(dashed line) in the saturated regime and deviates from that scaling law in the pre-saturate regime.}
      \label{fig:fastquench}
 \end{figure}

    Additional attention is placed on the fast quench dynamics. In Fig. \ref{fig:fastquench}(a), we collect the value of plateau $n_\text{su}$ as a function of the reverse of initial potential strength within very fast quench rate. A fitting for $1/2-n_\text{su}= \alpha/V_i$ yields the fitting parameter $\alpha \approx 0.425\pm 0.0004$. Furthermore, in Fig. \ref{fig:fastquench}(b), we show $(n_\text{su}-n_\text{ex})V_i$ as a function of $V_{i}^{2} \tau_Q$ by varying initial potential strength $V_i$. The curves with different $V_i$ share a universal scaling $(n_\text{su}-n_\text{ex})V_i\sim V_{i}^{2} \tau_Q$, which corresponds to Eq. (\ref{eq:saturate-d}) with the fitting parameter $K\approx 0.10\pm0.005$. In PSR, the turning point of SR and PSR, being proportional to the reciprocal of the squared initial potential strength, $\tau_Q^{S} \sim V_i^{-2}$, is reflected as an obvious change of the behavior of combined defects' density, $(n_\text{su}-n_\text{ex})V_i$, in Fig. \ref{fig:fastquench}(b). The scaling behavior of defect density is similar to that in the transverse Ising model, which corresponds to the similar critical behavior with $\nu,z = 1$.

    To sum up, the one-way quench dynamics reflect the two-plateau structure of the dynamical exponent (\ref{eq:dynamicalexponent}) according to the alternation of the KZ exponent in the KZ regime.

\subsection{Round-trip quench protocol}\label{Section:RT}
    Next, we analyze the influence of the dynamical exponent through the round-trip quench protocol, which is characterized by the interference effect of two critical dynamics \cite{Kou2022}. The interference effect occurs when the system undergoes multiple crossings of a transition point because of the dynamical phase. As one of the distinguishing features of this effect, the defect density as a function of the quench rate oscillates with a steady periodic $T_Q$. In this protocol, the system initially undergoes a transition from the potential strength $V=V_i$ in the localized phase to the turning point $V=V_{rt}$ in the critical localized phase by crossing the critical point with a quench time $\tau_Q$. Subsequently, the system drives back to the initial point with another quench time $\tau_Q'$. This round-trip protocol can be parameterized as
    \begin{equation}
    V \rightarrow V(t) = \left\{
    \begin{aligned}
    &V_{rt} - \frac{t}{\tau_Q}, t_i < t \leq 0, \\
    &V_{rt} + \frac{t}{\tau_Q^{'}}, 0< t < t_f,
    \end{aligned}
    \right.
    \end{equation}\label{eq:rtquench}
    where $V_{rt}$ represents the turning point in Fig. \ref{fig:spectrum}, and $\tau_Q$ and $\tau_Q^{'}$ are the quench times for the forward and backward processes, respectively, with a ratio $R = \tau_Q^{'}/\tau_Q$. We choose $t_i = (V_{rt}-V_{i})\tau_Q$, which corresponds to the initial deeply localized ground state, and $t_f = (V_{i}-V_{rt})\tau_Q^{'}$. The density of excitation probability caused by this procedure, denoted as $n_\text{ex}(t_f)$, can be numerically calculated using Eq. (\ref{eq:defects}).

\begin{figure}[t]
  \begin{center}
    		\includegraphics[width=3.2in,angle=0]{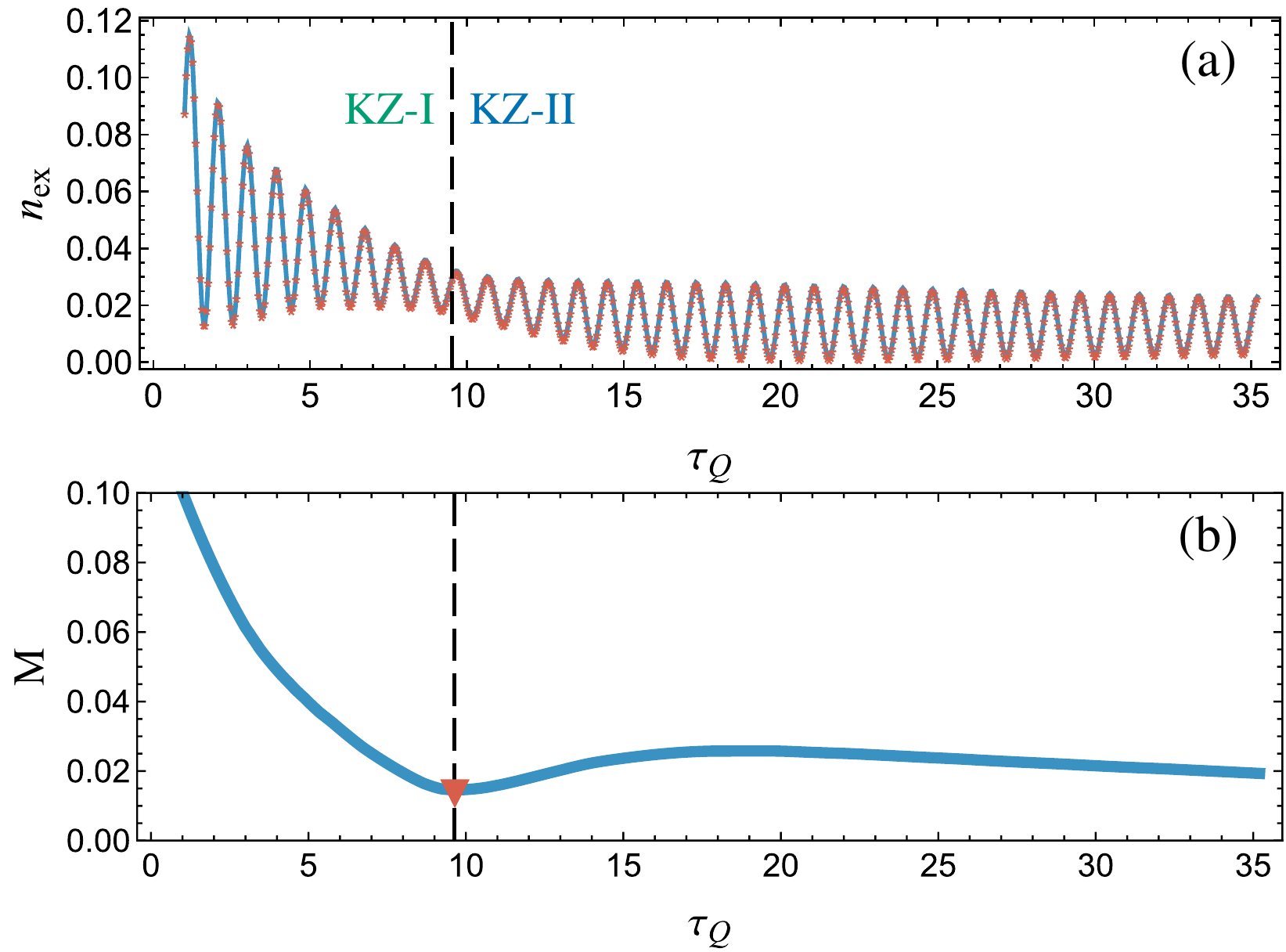}
  \end{center}
  \caption{The oscillatory defects and its amplitude as a function of quench time. (a) the final defects are obtained from the round-trip quench by adjusting the quench rate. The defects $n_\text{ex}$ oscillate with a period $T_Q\approx0.943$. (b) the amplitude as a function of quench time. The amplitude was obtained from the differences between interpolation functions of the peak and trough of the oscillatory defects. A minimum value of the amplitude is used to find the neck position $\tau_Q^\text{neck}$, which is well located at the turning point between two KZ sub-regimes, KZ-I and KZ-I.}
  \label{fig:interference001}
\end{figure}

\begin{figure}[t]
  \begin{center}
		\includegraphics[width=3.2in,angle=0]{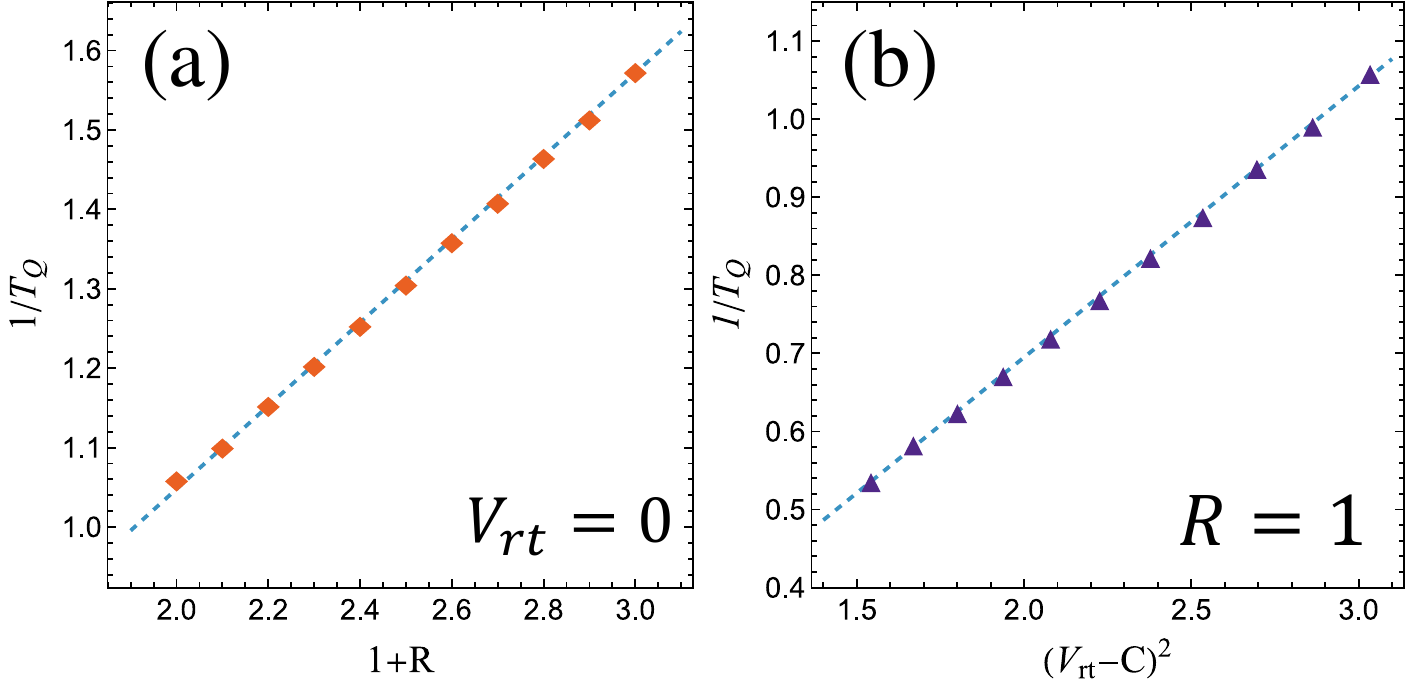}
  \end{center}
  \caption{The period of oscillatory defects obtained by varying to-and-fro quench time ratio $R$ and turning point $V_{rt}$. (a) $1/T_Q$ versus $1+R$ with $V_{rt}=1$. A linear model fitting of $1/T_Q = a (1+R)$ yields $a \approx 0.524\pm 0.0005$. (b) $1/T_Q$ versus $\left(V_{rt}-C\right)^2$ with $R=1$. A fitting of $1/T_Q = b \left(V_{rt}-C\right)^2 $ yields $b \approx 0.347 \pm 0.004$ and $C \approx 1.742\pm 0.009$.}
  \label{fig:period}
\end{figure}

\begin{figure}[t]
  \begin{center}
		\includegraphics[width=3.2in,angle=0]{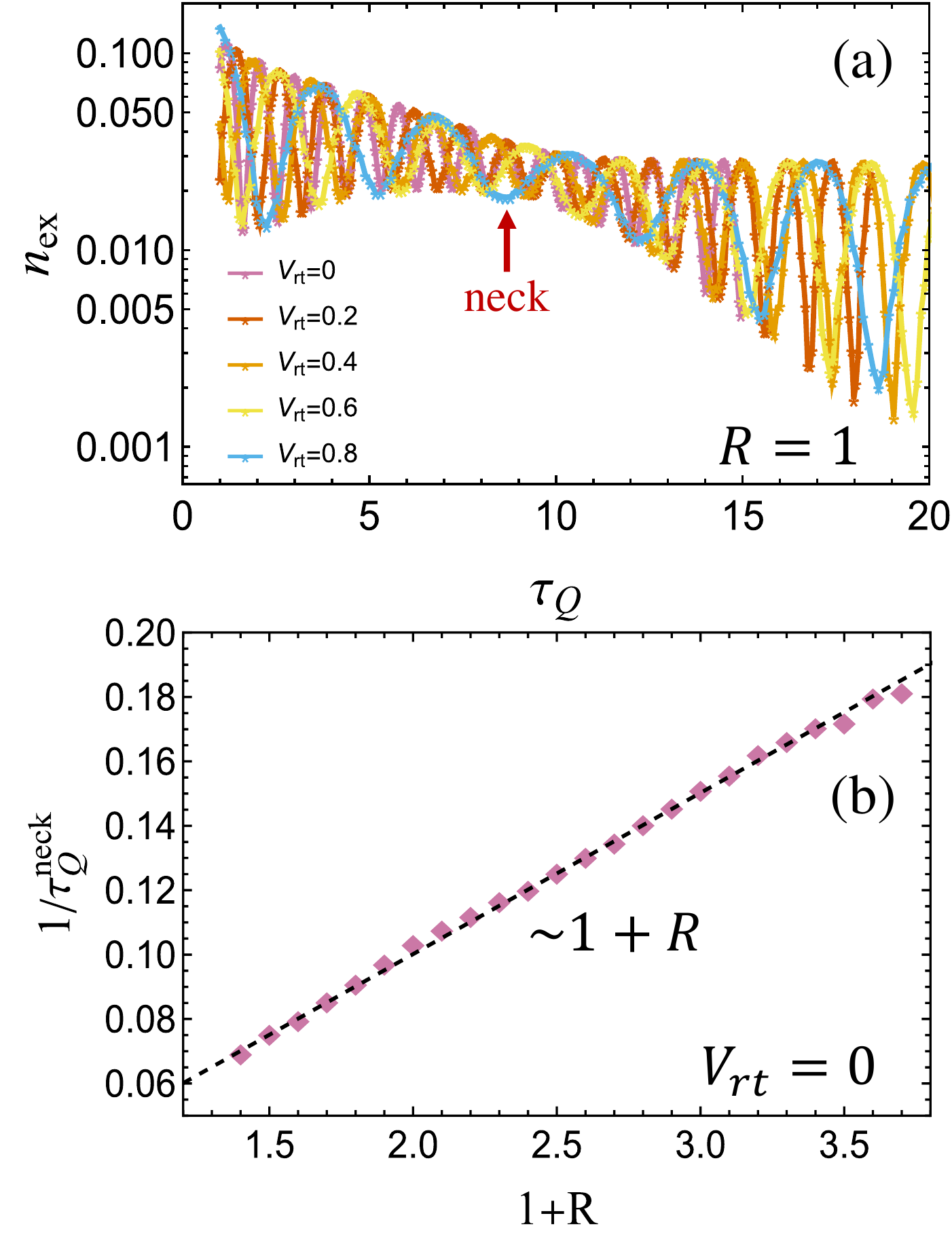}
  \end{center}
  \caption{(a) the density of defect as a function of the quench time $\tau_Q$ by setting the ratio of to-and-fro quench time $R=1$. The narrow neck is exposed by curves with different turning points $V_{rt}$. (b) The reciprocal of neck point $1/\tau_Q^\text{neck}$ versus $1+R$ with turning point $V_{rt}=0$. A linear fitting of $1/\tau_Q^\text{neck}=\lambda(1+R)$ yields $\lambda \approx 0.05 \pm 0.0001$.}
  \label{fig:neck}
\end{figure}

\subsubsection{Interference effect at $V_{rt}=0$ and $R = 1$}
    To observe the interference effect of defect density, we initially set the turning point $V_{rt} = 0$ and the ratio of to-and-fro quench times $R = 1$. In this protocol, the system starts in a deeply localized state with a potential strength of $V_i = 4$. Then, it undergoes a linear quench crossing the critical point to reach the turning point $V = V_{rt}$, and turning back to the initial potential strength passing through the critical point twice with the same quench rates.

    The final defect's density as a function of the quench time $\tau_Q$ is shown in Fig. \ref{fig:interference001}. After the round-trip protocol, the defect's density exhibits oscillatory behavior with a steady period $T_Q \approx 0.943$ because of the dynamical phase. Interestingly, a narrow neck appears in the oscillatory defect density after the round-trip protocol. To identify the position of the narrow neck, the amplitude $M$ is introduced by the differences between two interpolation functions from wave peaks and troughs (Fig. \ref{fig:interference001}(b)). The neck position $\tau_Q^\text{neck}$, defined as the minimum of the amplitude $M$, is well located between KZ-I and KZ-II. In this case, we suggest the narrow neck is a phenomenon ascribed as the changing dynamical exponent (\ref{eq:dynamicalexponent}). Notably, the interference effect is hard to be observed in the SR and PSR because the period $T_Q$ is larger than the scale of those regimes.

\subsubsection{Interference effect for varying $V_{rt}$ and $R$}
    Different turning points $V_{rt}$ and the quench times ratio $R$ are investigated for further investigating the interference effect. The oscillation periods, as functions of the to-and-fro quench times ratio $R$ and the turning point $V_{rt}$, are collected in Fig. \ref{fig:period}. In Fig. \ref{fig:period}(a), the turning point $V_{rt}$ is set to $0$, and the final defects' density is obtained by varying $R$. A linear fitting of $1/T_Q=a(1+R)$  yields the proportionality coefficient $a \approx 0.524 \pm 0.0005$. In Fig. \ref{fig:period} (b), with a fixed to-and-fro quench times ratio $R=1$, a fitting of $1/T_Q = b (V_{rt}-C)^2$ provides the proportionality coefficient $b \approx 0.347 \pm 0.004$ and the constant $C \approx 1.742 \pm 0.009$. These numerical results determine a relationship for the oscillation period:
     \begin{equation}
     T_Q \approx \frac{5.78 \pm0.09}{(1+R)(V_{rt}-C)^2},
     \end{equation}
    which is similar to the results in the transverse field Ising model \cite{Kou2022}.

    Furthermore, the neck position is also investigated by varying $V_{rt}$ and $R$. In Fig. \ref{fig:neck}(a), setting the quench time ratio at $R=1$, a distinct narrow neck is disclosed by varying the turning point $V_{rt}$. The position of this neck position is independent of $V_{rt}$, which reveals that the neck is irrelevant to the dynamical phase. In Fig. \ref{fig:neck}(b), we set the first linear ramps' quench time $\tau_Q$ and adjust the second one $\tau_Q^{'}$ to control the to-and-fro quench times ratio $R$, with $V_{rt}=0$. The neck position is found to be a function of the reciprocal of $1+R$, i.e. $\tau_Q^\text{neck}\sim(1+R)^{-1}$. The neck position with $R=1$ corresponds to the turning point between KZ-I and KZ-II (Fig. \ref{fig:interference001}(b)). An increased returning quench rate (associated with larger $R$) induces earlier entry into the KZ-II regime during the return process. This advancement shifts the neck position to occur at an earlier stage compared to the $R=1$ case, with reduced $R$ conversely delaying this event.

    In conclusion, the two-plateau structure of the dynamical exponent (\ref{eq:dynamicalexponent}) is reflected by a narrow neck in the oscillatory defect density after the round-trip quench protocol. Additionally, the position of narrow neck is well located between two KZ sub-regimes, which can be utilized to determine the turning point of KZ-I and KZ-II.

\subsection{Varying the model parameter $\Delta$}\label{Section:varyingdelta}
    Then we investigate the influence of the two-plateau structure (\ref{eq:dynamicalexponent}) by adjusting the system parameters $\Delta$. The final density of defects is collected as a function of quench time after one-way and round-trip quench protocols in the system with different the p-wave pairing amplitude $\Delta$. One should notice that the system across the extended-critical transition point will not get excitation because the energy gap is open at this transition point $V_{c_2} = \left|J-\Delta\right|$.

    \begin{figure}[t]
      \begin{center}
    		\includegraphics[width=3.2in,angle=0]{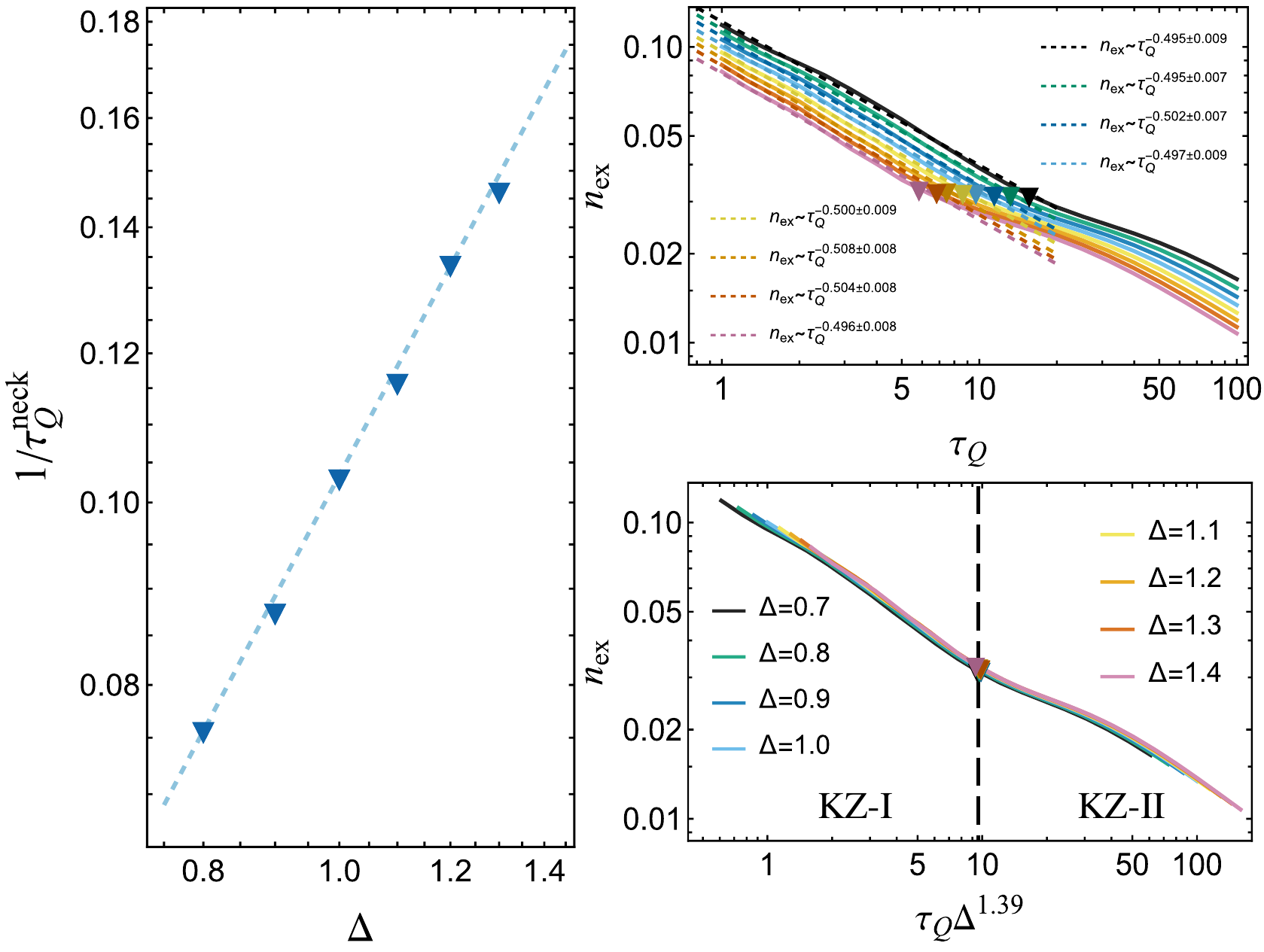}
      \end{center}
      \caption{Scaling of the neck position. (a) The reciprocal of the neck position $1/\tau_Q^\text{neck}$ as a function of the p-wave superconductive pairing amplitude $\Delta$. A linear model fitting of $\tau_Q^\text{neck} \sim \Delta^{-h}$ yields the power-law exponent $h=1.39\pm0.03$. (b) The defects after one-way quench $n_\text{ex}$ versus quench time $\tau_Q$ are collected by varying $\Delta$. For each $\Delta$, we show that the defect behaves as $n_\text{ex} \sim \tau_Q^{-0.5}$ for relevant fast quench rate (dash lines) at first and then deviates at the turning point $\tau_Q = \tau_Q^\text{neck}$. (c) $n_\text{ex}$  versus $\tau_Q \Delta^{1.39}$. The curves with different $\Delta$ collapse into a universal curve with a turning point of the KZ exponent.}
      \label{fig:varyingdelta}
\end{figure}

    In Fig. \ref{fig:varyingdelta}(a), the neck positions, extracted from the round-trip quench, as a function of p-wave superconductivity pairing amplitude $\Delta$ show a power-law behavior $\tau_Q^\text{neck}\sim\Delta^{-h}$ with $h=1.39\pm0.03$. In Fig. \ref{fig:varyingdelta}(b), we obtain the final excitation density populated after one-way quenching versus the quench time $\tau_Q$ by varying p-wave superconductive pairing amplitudes $\Delta$. The fits of $n_\text{ex} \sim \tau_Q^{-r}$ yields $r\approx0.5$ for each $\Delta$ within moderately fast quench rate, corresponding to the dash lines. The fitting is only applied for KZ-I because the power-law scaling of KZ-II occurs when there is a long quenching period. For different $\Delta$, neck positions are observed between the turning point of two sub-regimes with different KZ scaling behavior. By multiplying the scaled factor $\Delta^{1.39}$ by the quench time, different power-law scaling is collapsed to a universal curve in Fig. \ref{fig:varyingdelta}(c). The common neck position can divide the KZ regime into two sub-regimes, which shows that the influence the critical dynamics holds universally across all model parameters $\Delta$.

\section{Conclusion}
    In the gapless localized phase of the p-wave paired Aubry-André-Harper model, we have found that the dynamical exponent is continuously dependent on the potential strength. As the main feature, two plateaus have been identified in the dynamical exponent profile: one in the vicinity of the quantum phase transition point and another away from it. This two-plateau structure of the dynamical exponent turns out to influence the critical dynamics deeply. In the one-way quench protocol, we illustrate that the usual KZ regime consists of two sub-regimes with different KZ exponents as a consequence. While in the round-trip quench protocol, we demonstrate a narrow neck occurring in the oscillatory density of defects as an interference effect of two critical dynamics. The phenomenon of narrow neck is a direct reflection of the two-plateau structure of the dynamical exponent, which means such an interferometry provides a valuable method for probing the critical behavior of the quasiperiodic system. By adjusting the to-and-fro quench time ratio and model parameters, we show how the position of the neck can be used to quantitatively determine the turning point of the two KZ sub-regimes.

\section{ACKNOWLEDGMENTS}

We thank Jian-Song Pan, Yan He, B. Het\'enyi, L. Sanchez-Palencia, G. Roosz, and Uwe R. Fischer for insightful discussions.  This work is supported by NSFC under Grant No. 11074177.

\bibliographystyle{apsrev4-2}
\bibliography{citation}

\begin{thebibliography}{77}%
\makeatletter
\providecommand \@ifxundefined [1]{%
 \@ifx{#1\undefined}
}%
\providecommand \@ifnum [1]{%
 \ifnum #1\expandafter \@firstoftwo
 \else \expandafter \@secondoftwo
 \fi
}%
\providecommand \@ifx [1]{%
 \ifx #1\expandafter \@firstoftwo
 \else \expandafter \@secondoftwo
 \fi
}%
\providecommand \natexlab [1]{#1}%
\providecommand \enquote  [1]{``#1''}%
\providecommand \bibnamefont  [1]{#1}%
\providecommand \bibfnamefont [1]{#1}%
\providecommand \citenamefont [1]{#1}%
\providecommand \href@noop [0]{\@secondoftwo}%
\providecommand \href [0]{\begingroup \@sanitize@url \@href}%
\providecommand \@href[1]{\@@startlink{#1}\@@href}%
\providecommand \@@href[1]{\endgroup#1\@@endlink}%
\providecommand \@sanitize@url [0]{\catcode `\\12\catcode `\$12\catcode
  `\&12\catcode `\#12\catcode `\^12\catcode `\_12\catcode `\%12\relax}%
\providecommand \@@startlink[1]{}%
\providecommand \@@endlink[0]{}%
\providecommand \url  [0]{\begingroup\@sanitize@url \@url }%
\providecommand \@url [1]{\endgroup\@href {#1}{\urlprefix }}%
\providecommand \urlprefix  [0]{URL }%
\providecommand \Eprint [0]{\href }%
\providecommand \doibase [0]{https://doi.org/}%
\providecommand \selectlanguage [0]{\@gobble}%
\providecommand \bibinfo  [0]{\@secondoftwo}%
\providecommand \bibfield  [0]{\@secondoftwo}%
\providecommand \translation [1]{[#1]}%
\providecommand \BibitemOpen [0]{}%
\providecommand \bibitemStop [0]{}%
\providecommand \bibitemNoStop [0]{.\EOS\space}%
\providecommand \EOS [0]{\spacefactor3000\relax}%
\providecommand \BibitemShut  [1]{\csname bibitem#1\endcsname}%
\let\auto@bib@innerbib\@empty
\bibitem [{\citenamefont {Kibble}(1976)}]{Kibble1976}%
  \BibitemOpen
  \bibfield  {author} {\bibinfo {author} {\bibfnamefont {T.~W.~B.}\
  \bibnamefont {Kibble}},\ }\href {https://doi.org/10.1088/0305-4470/9/8/029}
  {\bibfield  {journal} {\bibinfo  {journal} {Journal of Physics A:
  Mathematical and General}\ }\textbf {\bibinfo {volume} {9}},\ \bibinfo
  {pages} {1387} (\bibinfo {year} {1976})}\BibitemShut {NoStop}%
\bibitem [{\citenamefont {Kibble}(1980)}]{Kibble1980}%
  \BibitemOpen
  \bibfield  {author} {\bibinfo {author} {\bibfnamefont {T.}~\bibnamefont
  {Kibble}},\ }\href
  {https://doi.org/https://doi.org/10.1016/0370-1573(80)90091-5} {\bibfield
  {journal} {\bibinfo  {journal} {Physics Reports}\ }\textbf {\bibinfo {volume}
  {67}},\ \bibinfo {pages} {183} (\bibinfo {year} {1980})}\BibitemShut
  {NoStop}%
\bibitem [{\citenamefont {Zurek}(1985)}]{Zurek1985}%
  \BibitemOpen
  \bibfield  {author} {\bibinfo {author} {\bibfnamefont {W.~H.}\ \bibnamefont
  {Zurek}},\ }\href {https://doi.org/10.1038/317505a0} {\bibfield  {journal}
  {\bibinfo  {journal} {Nature}\ }\textbf {\bibinfo {volume} {317}},\ \bibinfo
  {pages} {505} (\bibinfo {year} {1985})}\BibitemShut {NoStop}%
\bibitem [{\citenamefont {Dziarmaga}(2005)}]{Dziarmaga2005}%
  \BibitemOpen
  \bibfield  {author} {\bibinfo {author} {\bibfnamefont {J.}~\bibnamefont
  {Dziarmaga}},\ }\href {https://doi.org/10.1103/PhysRevLett.95.245701}
  {\bibfield  {journal} {\bibinfo  {journal} {Phys. Rev. Lett.}\ }\textbf
  {\bibinfo {volume} {95}},\ \bibinfo {pages} {245701} (\bibinfo {year}
  {2005})}\BibitemShut {NoStop}%
\bibitem [{\citenamefont {Sch\"utzhold}\ \emph {et~al.}(2006)\citenamefont
  {Sch\"utzhold}, \citenamefont {Uhlmann}, \citenamefont {Xu},\ and\
  \citenamefont {Fischer}}]{Fischer2006}%
  \BibitemOpen
  \bibfield  {author} {\bibinfo {author} {\bibfnamefont {R.}~\bibnamefont
  {Sch\"utzhold}}, \bibinfo {author} {\bibfnamefont {M.}~\bibnamefont
  {Uhlmann}}, \bibinfo {author} {\bibfnamefont {Y.}~\bibnamefont {Xu}},\ and\
  \bibinfo {author} {\bibfnamefont {U.~R.}\ \bibnamefont {Fischer}},\ }\href
  {https://doi.org/10.1103/PhysRevLett.97.200601} {\bibfield  {journal}
  {\bibinfo  {journal} {Phys. Rev. Lett.}\ }\textbf {\bibinfo {volume} {97}},\
  \bibinfo {pages} {200601} (\bibinfo {year} {2006})}\BibitemShut {NoStop}%
\bibitem [{\citenamefont {Uhlmann}\ \emph {et~al.}(2007)\citenamefont
  {Uhlmann}, \citenamefont {Sch\"utzhold},\ and\ \citenamefont
  {Fischer}}]{Fischer2007}%
  \BibitemOpen
  \bibfield  {author} {\bibinfo {author} {\bibfnamefont {M.}~\bibnamefont
  {Uhlmann}}, \bibinfo {author} {\bibfnamefont {R.}~\bibnamefont
  {Sch\"utzhold}},\ and\ \bibinfo {author} {\bibfnamefont {U.~R.}\ \bibnamefont
  {Fischer}},\ }\href {https://doi.org/10.1103/PhysRevLett.99.120407}
  {\bibfield  {journal} {\bibinfo  {journal} {Phys. Rev. Lett.}\ }\textbf
  {\bibinfo {volume} {99}},\ \bibinfo {pages} {120407} (\bibinfo {year}
  {2007})}\BibitemShut {NoStop}%
\bibitem [{\citenamefont {Fischer}\ \emph {et~al.}(2008)\citenamefont
  {Fischer}, \citenamefont {Sch\"utzhold},\ and\ \citenamefont
  {Uhlmann}}]{Fischer2008}%
  \BibitemOpen
  \bibfield  {author} {\bibinfo {author} {\bibfnamefont {U.~R.}\ \bibnamefont
  {Fischer}}, \bibinfo {author} {\bibfnamefont {R.}~\bibnamefont
  {Sch\"utzhold}},\ and\ \bibinfo {author} {\bibfnamefont {M.}~\bibnamefont
  {Uhlmann}},\ }\href {https://doi.org/10.1103/PhysRevA.77.043615} {\bibfield
  {journal} {\bibinfo  {journal} {Phys. Rev. A}\ }\textbf {\bibinfo {volume}
  {77}},\ \bibinfo {pages} {043615} (\bibinfo {year} {2008})}\BibitemShut
  {NoStop}%
\bibitem [{\citenamefont {Uhlmann}\ \emph {et~al.}(2010)\citenamefont
  {Uhlmann}, \citenamefont {Sch\"utzhold},\ and\ \citenamefont
  {Fischer}}]{Fischer2010}%
  \BibitemOpen
  \bibfield  {author} {\bibinfo {author} {\bibfnamefont {M.}~\bibnamefont
  {Uhlmann}}, \bibinfo {author} {\bibfnamefont {R.}~\bibnamefont
  {Sch\"utzhold}},\ and\ \bibinfo {author} {\bibfnamefont {U.~R.}\ \bibnamefont
  {Fischer}},\ }\href {https://doi.org/10.1103/PhysRevD.81.025017} {\bibfield
  {journal} {\bibinfo  {journal} {Phys. Rev. D}\ }\textbf {\bibinfo {volume}
  {81}},\ \bibinfo {pages} {025017} (\bibinfo {year} {2010})}\BibitemShut
  {NoStop}%
\bibitem [{\citenamefont {Dutta}\ \emph {et~al.}(2016)\citenamefont {Dutta},
  \citenamefont {Rahmani},\ and\ \citenamefont {del Campo}}]{Campol2016}%
  \BibitemOpen
  \bibfield  {author} {\bibinfo {author} {\bibfnamefont {A.}~\bibnamefont
  {Dutta}}, \bibinfo {author} {\bibfnamefont {A.}~\bibnamefont {Rahmani}},\
  and\ \bibinfo {author} {\bibfnamefont {A.}~\bibnamefont {del Campo}},\ }\href
  {https://doi.org/10.1103/PhysRevLett.117.080402} {\bibfield  {journal}
  {\bibinfo  {journal} {Phys. Rev. Lett.}\ }\textbf {\bibinfo {volume} {117}},\
  \bibinfo {pages} {080402} (\bibinfo {year} {2016})}\BibitemShut {NoStop}%
\bibitem [{\citenamefont {Keesling}\ \emph {et~al.}(2019)\citenamefont
  {Keesling}, \citenamefont {Omran}, \citenamefont {Levine}, \citenamefont
  {Bernien}, \citenamefont {Pichler}, \citenamefont {Choi}, \citenamefont
  {Samajdar}, \citenamefont {Schwartz}, \citenamefont {Silvi}, \citenamefont
  {Sachdev}, \citenamefont {Zoller}, \citenamefont {Endres}, \citenamefont
  {Greiner}, \citenamefont {Vuleti{\'{c}}},\ and\ \citenamefont
  {Lukin}}]{Keesling2019}%
  \BibitemOpen
  \bibfield  {author} {\bibinfo {author} {\bibfnamefont {A.}~\bibnamefont
  {Keesling}}, \bibinfo {author} {\bibfnamefont {A.}~\bibnamefont {Omran}},
  \bibinfo {author} {\bibfnamefont {H.}~\bibnamefont {Levine}}, \bibinfo
  {author} {\bibfnamefont {H.}~\bibnamefont {Bernien}}, \bibinfo {author}
  {\bibfnamefont {H.}~\bibnamefont {Pichler}}, \bibinfo {author} {\bibfnamefont
  {S.}~\bibnamefont {Choi}}, \bibinfo {author} {\bibfnamefont {R.}~\bibnamefont
  {Samajdar}}, \bibinfo {author} {\bibfnamefont {S.}~\bibnamefont {Schwartz}},
  \bibinfo {author} {\bibfnamefont {P.}~\bibnamefont {Silvi}}, \bibinfo
  {author} {\bibfnamefont {S.}~\bibnamefont {Sachdev}}, \bibinfo {author}
  {\bibfnamefont {P.}~\bibnamefont {Zoller}}, \bibinfo {author} {\bibfnamefont
  {M.}~\bibnamefont {Endres}}, \bibinfo {author} {\bibfnamefont
  {M.}~\bibnamefont {Greiner}}, \bibinfo {author} {\bibfnamefont
  {V.}~\bibnamefont {Vuleti{\'{c}}}},\ and\ \bibinfo {author} {\bibfnamefont
  {M.~D.}\ \bibnamefont {Lukin}},\ }\href
  {https://doi.org/10.1038/s41586-019-1070-1} {\bibfield  {journal} {\bibinfo
  {journal} {Nature}\ }\textbf {\bibinfo {volume} {568}},\ \bibinfo {pages}
  {207} (\bibinfo {year} {2019})}\BibitemShut {NoStop}%
\bibitem [{\citenamefont {Reichhardt}\ \emph {et~al.}(2022)\citenamefont
  {Reichhardt}, \citenamefont {del Campo},\ and\ \citenamefont
  {Reichhardt}}]{Campol2022}%
  \BibitemOpen
  \bibfield  {author} {\bibinfo {author} {\bibfnamefont {C.~J.~O.}\
  \bibnamefont {Reichhardt}}, \bibinfo {author} {\bibfnamefont
  {A.}~\bibnamefont {del Campo}},\ and\ \bibinfo {author} {\bibfnamefont
  {C.}~\bibnamefont {Reichhardt}},\ }\href
  {https://doi.org/10.1038/s42005-022-00952-w} {\bibfield  {journal} {\bibinfo
  {journal} {Communications Physics}\ }\textbf {\bibinfo {volume} {5}},\
  \bibinfo {pages} {173} (\bibinfo {year} {2022})}\BibitemShut {NoStop}%
\bibitem [{\citenamefont {B{\'a}csi}\ and\ \citenamefont
  {D{\'o}ra}(2023)}]{Bacsi2023}%
  \BibitemOpen
  \bibfield  {author} {\bibinfo {author} {\bibfnamefont {{\'A}.}~\bibnamefont
  {B{\'a}csi}}\ and\ \bibinfo {author} {\bibfnamefont {B.}~\bibnamefont
  {D{\'o}ra}},\ }\href {https://doi.org/10.1038/s41598-023-30840-4} {\bibfield
  {journal} {\bibinfo  {journal} {Scientific Reports}\ }\textbf {\bibinfo
  {volume} {13}},\ \bibinfo {pages} {4034} (\bibinfo {year}
  {2023})}\BibitemShut {NoStop}%
\bibitem [{\citenamefont {Het\'enyi}(2024)}]{Hetenyi2024}%
  \BibitemOpen
  \bibfield  {author} {\bibinfo {author} {\bibfnamefont {B.}~\bibnamefont
  {Het\'enyi}},\ }\href {https://doi.org/10.1103/PhysRevB.110.125124}
  {\bibfield  {journal} {\bibinfo  {journal} {Phys. Rev. B}\ }\textbf {\bibinfo
  {volume} {110}},\ \bibinfo {pages} {125124} (\bibinfo {year}
  {2024})}\BibitemShut {NoStop}%
\bibitem [{\citenamefont {Kou}\ and\ \citenamefont {Li}(2022)}]{Kou2022}%
  \BibitemOpen
  \bibfield  {author} {\bibinfo {author} {\bibfnamefont {H.-C.}\ \bibnamefont
  {Kou}}\ and\ \bibinfo {author} {\bibfnamefont {P.}~\bibnamefont {Li}},\
  }\href {https://doi.org/10.1103/PhysRevB.106.184301} {\bibfield  {journal}
  {\bibinfo  {journal} {Phys. Rev. B}\ }\textbf {\bibinfo {volume} {106}},\
  \bibinfo {pages} {184301} (\bibinfo {year} {2022})}\BibitemShut {NoStop}%
\bibitem [{\citenamefont {del Campo}\ \emph {et~al.}(2010)\citenamefont {del
  Campo}, \citenamefont {De~Chiara}, \citenamefont {Morigi}, \citenamefont
  {Plenio},\ and\ \citenamefont {Retzker}}]{Campo2010}%
  \BibitemOpen
  \bibfield  {author} {\bibinfo {author} {\bibfnamefont {A.}~\bibnamefont {del
  Campo}}, \bibinfo {author} {\bibfnamefont {G.}~\bibnamefont {De~Chiara}},
  \bibinfo {author} {\bibfnamefont {G.}~\bibnamefont {Morigi}}, \bibinfo
  {author} {\bibfnamefont {M.~B.}\ \bibnamefont {Plenio}},\ and\ \bibinfo
  {author} {\bibfnamefont {A.}~\bibnamefont {Retzker}},\ }\href
  {https://doi.org/10.1103/PhysRevLett.105.075701} {\bibfield  {journal}
  {\bibinfo  {journal} {Phys. Rev. Lett.}\ }\textbf {\bibinfo {volume} {105}},\
  \bibinfo {pages} {075701} (\bibinfo {year} {2010})}\BibitemShut {NoStop}%
\bibitem [{\citenamefont {Griffin}\ \emph {et~al.}(2012)\citenamefont
  {Griffin}, \citenamefont {Lilienblum}, \citenamefont {Delaney}, \citenamefont
  {Kumagai}, \citenamefont {Fiebig},\ and\ \citenamefont
  {Spaldin}}]{Griffin2012}%
  \BibitemOpen
  \bibfield  {author} {\bibinfo {author} {\bibfnamefont {S.~M.}\ \bibnamefont
  {Griffin}}, \bibinfo {author} {\bibfnamefont {M.}~\bibnamefont {Lilienblum}},
  \bibinfo {author} {\bibfnamefont {K.~T.}\ \bibnamefont {Delaney}}, \bibinfo
  {author} {\bibfnamefont {Y.}~\bibnamefont {Kumagai}}, \bibinfo {author}
  {\bibfnamefont {M.}~\bibnamefont {Fiebig}},\ and\ \bibinfo {author}
  {\bibfnamefont {N.~A.}\ \bibnamefont {Spaldin}},\ }\href
  {https://doi.org/10.1103/PhysRevX.2.041022} {\bibfield  {journal} {\bibinfo
  {journal} {Phys. Rev. X}\ }\textbf {\bibinfo {volume} {2}},\ \bibinfo {pages}
  {041022} (\bibinfo {year} {2012})}\BibitemShut {NoStop}%
\bibitem [{\citenamefont {Sonner}\ \emph {et~al.}(2015)\citenamefont {Sonner},
  \citenamefont {del Campo},\ and\ \citenamefont {Zurek}}]{Sonner2015}%
  \BibitemOpen
  \bibfield  {author} {\bibinfo {author} {\bibfnamefont {J.}~\bibnamefont
  {Sonner}}, \bibinfo {author} {\bibfnamefont {A.}~\bibnamefont {del Campo}},\
  and\ \bibinfo {author} {\bibfnamefont {W.~H.}\ \bibnamefont {Zurek}},\
  }\href@noop {} {\bibfield  {journal} {\bibinfo  {journal} {Nature
  Communications}\ }\textbf {\bibinfo {volume} {6}},\ \bibinfo {pages} {7406}
  (\bibinfo {year} {2015})}\BibitemShut {NoStop}%
\bibitem [{\citenamefont {Chesler}\ \emph {et~al.}(2015)\citenamefont
  {Chesler}, \citenamefont {Garc\'{\i}a-Garc\'{\i}a},\ and\ \citenamefont
  {Liu}}]{Chesler2015}%
  \BibitemOpen
  \bibfield  {author} {\bibinfo {author} {\bibfnamefont {P.~M.}\ \bibnamefont
  {Chesler}}, \bibinfo {author} {\bibfnamefont {A.~M.}\ \bibnamefont
  {Garc\'{\i}a-Garc\'{\i}a}},\ and\ \bibinfo {author} {\bibfnamefont
  {H.}~\bibnamefont {Liu}},\ }\href {https://doi.org/10.1103/PhysRevX.5.021015}
  {\bibfield  {journal} {\bibinfo  {journal} {Phys. Rev. X}\ }\textbf {\bibinfo
  {volume} {5}},\ \bibinfo {pages} {021015} (\bibinfo {year}
  {2015})}\BibitemShut {NoStop}%
\bibitem [{\citenamefont {G\'omez-Ruiz}\ and\ \citenamefont {del
  Campo}(2019)}]{Campo2019}%
  \BibitemOpen
  \bibfield  {author} {\bibinfo {author} {\bibfnamefont {F.~J.}\ \bibnamefont
  {G\'omez-Ruiz}}\ and\ \bibinfo {author} {\bibfnamefont {A.}~\bibnamefont {del
  Campo}},\ }\href {https://doi.org/10.1103/PhysRevLett.122.080604} {\bibfield
  {journal} {\bibinfo  {journal} {Phys. Rev. Lett.}\ }\textbf {\bibinfo
  {volume} {122}},\ \bibinfo {pages} {080604} (\bibinfo {year}
  {2019})}\BibitemShut {NoStop}%
\bibitem [{\citenamefont {Fei}\ and\ \citenamefont {Sun}(2021)}]{Sun2021}%
  \BibitemOpen
  \bibfield  {author} {\bibinfo {author} {\bibfnamefont {Z.}~\bibnamefont
  {Fei}}\ and\ \bibinfo {author} {\bibfnamefont {C.~P.}\ \bibnamefont {Sun}},\
  }\href {https://doi.org/10.1103/PhysRevB.103.144204} {\bibfield  {journal}
  {\bibinfo  {journal} {Phys. Rev. B}\ }\textbf {\bibinfo {volume} {103}},\
  \bibinfo {pages} {144204} (\bibinfo {year} {2021})}\BibitemShut {NoStop}%
\bibitem [{\citenamefont {Zeng}\ \emph {et~al.}(2023)\citenamefont {Zeng},
  \citenamefont {Xia},\ and\ \citenamefont {del Campo}}]{Campo2023PRL}%
  \BibitemOpen
  \bibfield  {author} {\bibinfo {author} {\bibfnamefont {H.-B.}\ \bibnamefont
  {Zeng}}, \bibinfo {author} {\bibfnamefont {C.-Y.}\ \bibnamefont {Xia}},\ and\
  \bibinfo {author} {\bibfnamefont {A.}~\bibnamefont {del Campo}},\ }\href
  {https://doi.org/10.1103/PhysRevLett.130.060402} {\bibfield  {journal}
  {\bibinfo  {journal} {Phys. Rev. Lett.}\ }\textbf {\bibinfo {volume} {130}},\
  \bibinfo {pages} {060402} (\bibinfo {year} {2023})}\BibitemShut {NoStop}%
\bibitem [{\citenamefont {Yang}\ \emph {et~al.}(2023)\citenamefont {Yang},
  \citenamefont {Tsubota}, \citenamefont {del Campo},\ and\ \citenamefont
  {Zeng}}]{Campo2023PRB}%
  \BibitemOpen
  \bibfield  {author} {\bibinfo {author} {\bibfnamefont {W.-C.}\ \bibnamefont
  {Yang}}, \bibinfo {author} {\bibfnamefont {M.}~\bibnamefont {Tsubota}},
  \bibinfo {author} {\bibfnamefont {A.}~\bibnamefont {del Campo}},\ and\
  \bibinfo {author} {\bibfnamefont {H.-B.}\ \bibnamefont {Zeng}},\ }\href
  {https://doi.org/10.1103/PhysRevB.108.174518} {\bibfield  {journal} {\bibinfo
   {journal} {Phys. Rev. B}\ }\textbf {\bibinfo {volume} {108}},\ \bibinfo
  {pages} {174518} (\bibinfo {year} {2023})}\BibitemShut {NoStop}%
\bibitem [{\citenamefont {Xia}\ \emph {et~al.}(2023)\citenamefont {Xia},
  \citenamefont {Zeng}, \citenamefont {Chen},\ and\ \citenamefont {del
  Campo}}]{Campo2023PRD}%
  \BibitemOpen
  \bibfield  {author} {\bibinfo {author} {\bibfnamefont {C.-Y.}\ \bibnamefont
  {Xia}}, \bibinfo {author} {\bibfnamefont {H.-B.}\ \bibnamefont {Zeng}},
  \bibinfo {author} {\bibfnamefont {C.-M.}\ \bibnamefont {Chen}},\ and\
  \bibinfo {author} {\bibfnamefont {A.}~\bibnamefont {del Campo}},\ }\href
  {https://doi.org/10.1103/PhysRevD.108.026017} {\bibfield  {journal} {\bibinfo
   {journal} {Phys. Rev. D}\ }\textbf {\bibinfo {volume} {108}},\ \bibinfo
  {pages} {026017} (\bibinfo {year} {2023})}\BibitemShut {NoStop}%
\bibitem [{\citenamefont {Kou}\ and\ \citenamefont {Li}(2023)}]{Kou2023}%
  \BibitemOpen
  \bibfield  {author} {\bibinfo {author} {\bibfnamefont {H.-C.}\ \bibnamefont
  {Kou}}\ and\ \bibinfo {author} {\bibfnamefont {P.}~\bibnamefont {Li}},\
  }\href {https://doi.org/10.1103/PhysRevB.108.214307} {\bibfield  {journal}
  {\bibinfo  {journal} {Phys. Rev. B}\ }\textbf {\bibinfo {volume} {108}},\
  \bibinfo {pages} {214307} (\bibinfo {year} {2023})}\BibitemShut {NoStop}%
\bibitem [{\citenamefont {Anderson}(1958)}]{Anderson1958}%
  \BibitemOpen
  \bibfield  {author} {\bibinfo {author} {\bibfnamefont {P.~W.}\ \bibnamefont
  {Anderson}},\ }\href {https://doi.org/10.1103/PhysRev.109.1492} {\bibfield
  {journal} {\bibinfo  {journal} {Phys. Rev.}\ }\textbf {\bibinfo {volume}
  {109}},\ \bibinfo {pages} {1492} (\bibinfo {year} {1958})}\BibitemShut
  {NoStop}%
\bibitem [{\citenamefont {Evers}\ and\ \citenamefont
  {Mirlin}(2008)}]{Evers2008}%
  \BibitemOpen
  \bibfield  {author} {\bibinfo {author} {\bibfnamefont {F.}~\bibnamefont
  {Evers}}\ and\ \bibinfo {author} {\bibfnamefont {A.~D.}\ \bibnamefont
  {Mirlin}},\ }\href {https://doi.org/10.1103/RevModPhys.80.1355} {\bibfield
  {journal} {\bibinfo  {journal} {Rev. Mod. Phys.}\ }\textbf {\bibinfo {volume}
  {80}},\ \bibinfo {pages} {1355} (\bibinfo {year} {2008})}\BibitemShut
  {NoStop}%
\bibitem [{\citenamefont {St\"orzer}\ \emph {et~al.}(2006)\citenamefont
  {St\"orzer}, \citenamefont {Gross}, \citenamefont {Aegerter},\ and\
  \citenamefont {Maret}}]{Martin2006}%
  \BibitemOpen
  \bibfield  {author} {\bibinfo {author} {\bibfnamefont {M.}~\bibnamefont
  {St\"orzer}}, \bibinfo {author} {\bibfnamefont {P.}~\bibnamefont {Gross}},
  \bibinfo {author} {\bibfnamefont {C.~M.}\ \bibnamefont {Aegerter}},\ and\
  \bibinfo {author} {\bibfnamefont {G.}~\bibnamefont {Maret}},\ }\href
  {https://doi.org/10.1103/PhysRevLett.96.063904} {\bibfield  {journal}
  {\bibinfo  {journal} {Phys. Rev. Lett.}\ }\textbf {\bibinfo {volume} {96}},\
  \bibinfo {pages} {063904} (\bibinfo {year} {2006})}\BibitemShut {NoStop}%
\bibitem [{\citenamefont {Pixley}\ \emph {et~al.}(2015)\citenamefont {Pixley},
  \citenamefont {Goswami},\ and\ \citenamefont {Das~Sarma}}]{Sarma2015}%
  \BibitemOpen
  \bibfield  {author} {\bibinfo {author} {\bibfnamefont {J.~H.}\ \bibnamefont
  {Pixley}}, \bibinfo {author} {\bibfnamefont {P.}~\bibnamefont {Goswami}},\
  and\ \bibinfo {author} {\bibfnamefont {S.}~\bibnamefont {Das~Sarma}},\ }\href
  {https://doi.org/10.1103/PhysRevLett.115.076601} {\bibfield  {journal}
  {\bibinfo  {journal} {Phys. Rev. Lett.}\ }\textbf {\bibinfo {volume} {115}},\
  \bibinfo {pages} {076601} (\bibinfo {year} {2015})}\BibitemShut {NoStop}%
\bibitem [{\citenamefont {Semeghini}\ \emph {et~al.}(2015)\citenamefont
  {Semeghini}, \citenamefont {Landini}, \citenamefont {Castilho}, \citenamefont
  {Roy}, \citenamefont {Spagnolli}, \citenamefont {Trenkwalder}, \citenamefont
  {Fattori}, \citenamefont {Inguscio},\ and\ \citenamefont
  {Modugno}}]{Semeghini2015}%
  \BibitemOpen
  \bibfield  {author} {\bibinfo {author} {\bibfnamefont {G.}~\bibnamefont
  {Semeghini}}, \bibinfo {author} {\bibfnamefont {M.}~\bibnamefont {Landini}},
  \bibinfo {author} {\bibfnamefont {P.}~\bibnamefont {Castilho}}, \bibinfo
  {author} {\bibfnamefont {S.}~\bibnamefont {Roy}}, \bibinfo {author}
  {\bibfnamefont {G.}~\bibnamefont {Spagnolli}}, \bibinfo {author}
  {\bibfnamefont {A.}~\bibnamefont {Trenkwalder}}, \bibinfo {author}
  {\bibfnamefont {M.}~\bibnamefont {Fattori}}, \bibinfo {author} {\bibfnamefont
  {M.}~\bibnamefont {Inguscio}},\ and\ \bibinfo {author} {\bibfnamefont
  {G.}~\bibnamefont {Modugno}},\ }\href@noop {} {\bibfield  {journal} {\bibinfo
   {journal} {Nature Physics}\ }\textbf {\bibinfo {volume} {11}},\ \bibinfo
  {pages} {554} (\bibinfo {year} {2015})}\BibitemShut {NoStop}%
\bibitem [{\citenamefont {Delande}\ \emph {et~al.}(2017)\citenamefont
  {Delande}, \citenamefont {Morales-Molina},\ and\ \citenamefont
  {Sacha}}]{Delande2017}%
  \BibitemOpen
  \bibfield  {author} {\bibinfo {author} {\bibfnamefont {D.}~\bibnamefont
  {Delande}}, \bibinfo {author} {\bibfnamefont {L.}~\bibnamefont
  {Morales-Molina}},\ and\ \bibinfo {author} {\bibfnamefont {K.}~\bibnamefont
  {Sacha}},\ }\href {https://doi.org/10.1103/PhysRevLett.119.230404} {\bibfield
   {journal} {\bibinfo  {journal} {Phys. Rev. Lett.}\ }\textbf {\bibinfo
  {volume} {119}},\ \bibinfo {pages} {230404} (\bibinfo {year}
  {2017})}\BibitemShut {NoStop}%
\bibitem [{\citenamefont {Het\'enyi}\ \emph {et~al.}(2021)\citenamefont
  {Het\'enyi}, \citenamefont {Parlak},\ and\ \citenamefont
  {Yahyavi}}]{Hetenyi2021}%
  \BibitemOpen
  \bibfield  {author} {\bibinfo {author} {\bibfnamefont {B.}~\bibnamefont
  {Het\'enyi}}, \bibinfo {author} {\bibfnamefont {S.~m.~c.}\ \bibnamefont
  {Parlak}},\ and\ \bibinfo {author} {\bibfnamefont {M.}~\bibnamefont
  {Yahyavi}},\ }\href {https://doi.org/10.1103/PhysRevB.104.214207} {\bibfield
  {journal} {\bibinfo  {journal} {Phys. Rev. B}\ }\textbf {\bibinfo {volume}
  {104}},\ \bibinfo {pages} {214207} (\bibinfo {year} {2021})}\BibitemShut
  {NoStop}%
\bibitem [{\citenamefont {Wang}\ and\ \citenamefont {Wang}(2023)}]{Wang2023}%
  \BibitemOpen
  \bibfield  {author} {\bibinfo {author} {\bibfnamefont {C.}~\bibnamefont
  {Wang}}\ and\ \bibinfo {author} {\bibfnamefont {X.~R.}\ \bibnamefont
  {Wang}},\ }\href {https://doi.org/10.1103/PhysRevB.107.024202} {\bibfield
  {journal} {\bibinfo  {journal} {Phys. Rev. B}\ }\textbf {\bibinfo {volume}
  {107}},\ \bibinfo {pages} {024202} (\bibinfo {year} {2023})}\BibitemShut
  {NoStop}%
\bibitem [{\citenamefont {Chen}\ \emph {et~al.}(2024)\citenamefont {Chen},
  \citenamefont {Maciejko},\ and\ \citenamefont {Boettcher}}]{Chen2024}%
  \BibitemOpen
  \bibfield  {author} {\bibinfo {author} {\bibfnamefont {A.}~\bibnamefont
  {Chen}}, \bibinfo {author} {\bibfnamefont {J.}~\bibnamefont {Maciejko}},\
  and\ \bibinfo {author} {\bibfnamefont {I.}~\bibnamefont {Boettcher}},\ }\href
  {https://doi.org/10.1103/PhysRevLett.133.066101} {\bibfield  {journal}
  {\bibinfo  {journal} {Phys. Rev. Lett.}\ }\textbf {\bibinfo {volume} {133}},\
  \bibinfo {pages} {066101} (\bibinfo {year} {2024})}\BibitemShut {NoStop}%
\bibitem [{\citenamefont {Aubry}\ and\ \citenamefont
  {André}(1980)}]{Aubry1980}%
  \BibitemOpen
  \bibfield  {author} {\bibinfo {author} {\bibfnamefont {S.}~\bibnamefont
  {Aubry}}\ and\ \bibinfo {author} {\bibfnamefont {G.}~\bibnamefont {André}},\
  }\href@noop {} {\bibfield  {journal} {\bibinfo  {journal} {Proceedings, VIII
  International Colloquium on Group-Theoretical Methods in Physics}\ }\textbf
  {\bibinfo {volume} {3}} (\bibinfo {year} {1980})}\BibitemShut {NoStop}%
\bibitem [{\citenamefont {Harper}(1955)}]{Harper1955}%
  \BibitemOpen
  \bibfield  {author} {\bibinfo {author} {\bibfnamefont {P.~G.}\ \bibnamefont
  {Harper}},\ }\href {https://doi.org/10.1088/0370-1298/68/10/304} {\bibfield
  {journal} {\bibinfo  {journal} {Proceedings of the Physical Society. Section
  A}\ }\textbf {\bibinfo {volume} {68}},\ \bibinfo {pages} {874} (\bibinfo
  {year} {1955})}\BibitemShut {NoStop}%
\bibitem [{\citenamefont {Luck}\ and\ \citenamefont
  {Nieuwenhuizen}(1986)}]{Luck1986}%
  \BibitemOpen
  \bibfield  {author} {\bibinfo {author} {\bibfnamefont {J.~M.}\ \bibnamefont
  {Luck}}\ and\ \bibinfo {author} {\bibfnamefont {T.~M.}\ \bibnamefont
  {Nieuwenhuizen}},\ }\href {https://doi.org/10.1209/0295-5075/2/4/001}
  {\bibfield  {journal} {\bibinfo  {journal} {Europhysics Letters}\ }\textbf
  {\bibinfo {volume} {2}},\ \bibinfo {pages} {257} (\bibinfo {year}
  {1986})}\BibitemShut {NoStop}%
\bibitem [{\citenamefont {Doria}\ and\ \citenamefont
  {Satija}(1988)}]{Doria1988}%
  \BibitemOpen
  \bibfield  {author} {\bibinfo {author} {\bibfnamefont {M.~M.}\ \bibnamefont
  {Doria}}\ and\ \bibinfo {author} {\bibfnamefont {I.~I.}\ \bibnamefont
  {Satija}},\ }\href {https://doi.org/10.1103/PhysRevLett.60.444} {\bibfield
  {journal} {\bibinfo  {journal} {Phys. Rev. Lett.}\ }\textbf {\bibinfo
  {volume} {60}},\ \bibinfo {pages} {444} (\bibinfo {year} {1988})}\BibitemShut
  {NoStop}%
\bibitem [{\citenamefont {Doria}\ \emph {et~al.}(1989)\citenamefont {Doria},
  \citenamefont {Nori},\ and\ \citenamefont {Satija}}]{Doria1989}%
  \BibitemOpen
  \bibfield  {author} {\bibinfo {author} {\bibfnamefont {M.~M.}\ \bibnamefont
  {Doria}}, \bibinfo {author} {\bibfnamefont {F.}~\bibnamefont {Nori}},\ and\
  \bibinfo {author} {\bibfnamefont {I.~I.}\ \bibnamefont {Satija}},\ }\href
  {https://doi.org/10.1103/PhysRevB.39.6802} {\bibfield  {journal} {\bibinfo
  {journal} {Phys. Rev. B}\ }\textbf {\bibinfo {volume} {39}},\ \bibinfo
  {pages} {6802} (\bibinfo {year} {1989})}\BibitemShut {NoStop}%
\bibitem [{\citenamefont {Benza}(1989)}]{Benza1989}%
  \BibitemOpen
  \bibfield  {author} {\bibinfo {author} {\bibfnamefont {V.~G.}\ \bibnamefont
  {Benza}},\ }\href {https://doi.org/10.1209/0295-5075/8/4/004} {\bibfield
  {journal} {\bibinfo  {journal} {Europhysics Letters}\ }\textbf {\bibinfo
  {volume} {8}},\ \bibinfo {pages} {321} (\bibinfo {year} {1989})}\BibitemShut
  {NoStop}%
\bibitem [{\citenamefont {Satija}\ and\ \citenamefont
  {Doria}(1989)}]{Satija1989}%
  \BibitemOpen
  \bibfield  {author} {\bibinfo {author} {\bibfnamefont {I.~I.}\ \bibnamefont
  {Satija}}\ and\ \bibinfo {author} {\bibfnamefont {M.~M.}\ \bibnamefont
  {Doria}},\ }\href {https://doi.org/10.1103/PhysRevB.39.9757} {\bibfield
  {journal} {\bibinfo  {journal} {Phys. Rev. B}\ }\textbf {\bibinfo {volume}
  {39}},\ \bibinfo {pages} {9757} (\bibinfo {year} {1989})}\BibitemShut
  {NoStop}%
\bibitem [{\citenamefont {Luck}(1993)}]{Luck1993}%
  \BibitemOpen
  \bibfield  {author} {\bibinfo {author} {\bibfnamefont {J.~M.}\ \bibnamefont
  {Luck}},\ }\href@noop {} {\bibfield  {journal} {\bibinfo  {journal} {Journal
  of Statistical Physics}\ }\textbf {\bibinfo {volume} {72}},\ \bibinfo {pages}
  {417} (\bibinfo {year} {1993})}\BibitemShut {NoStop}%
\bibitem [{\citenamefont {Hermisson}\ \emph {et~al.}(1997)\citenamefont
  {Hermisson}, \citenamefont {Grimm},\ and\ \citenamefont
  {Baake}}]{Joachim1997}%
  \BibitemOpen
  \bibfield  {author} {\bibinfo {author} {\bibfnamefont {J.}~\bibnamefont
  {Hermisson}}, \bibinfo {author} {\bibfnamefont {U.}~\bibnamefont {Grimm}},\
  and\ \bibinfo {author} {\bibfnamefont {M.}~\bibnamefont {Baake}},\ }\href
  {https://doi.org/10.1088/0305-4470/30/21/009} {\bibfield  {journal} {\bibinfo
   {journal} {Journal of Physics A: Mathematical and General}\ }\textbf
  {\bibinfo {volume} {30}},\ \bibinfo {pages} {7315} (\bibinfo {year}
  {1997})}\BibitemShut {NoStop}%
\bibitem [{\citenamefont {Hermisson}\ and\ \citenamefont
  {Grimm}(1998)}]{Hermisson1998}%
  \BibitemOpen
  \bibfield  {author} {\bibinfo {author} {\bibfnamefont {J.}~\bibnamefont
  {Hermisson}}\ and\ \bibinfo {author} {\bibfnamefont {U.}~\bibnamefont
  {Grimm}},\ }\href {https://doi.org/10.1103/PhysRevB.57.R673} {\bibfield
  {journal} {\bibinfo  {journal} {Phys. Rev. B}\ }\textbf {\bibinfo {volume}
  {57}},\ \bibinfo {pages} {R673} (\bibinfo {year} {1998})}\BibitemShut
  {NoStop}%
\bibitem [{\citenamefont {Hermisson}(2000)}]{Hermisson2000}%
  \BibitemOpen
  \bibfield  {author} {\bibinfo {author} {\bibfnamefont {J.}~\bibnamefont
  {Hermisson}},\ }\href {https://doi.org/10.1088/0305-4470/33/1/304} {\bibfield
   {journal} {\bibinfo  {journal} {Journal of Physics A: Mathematical and
  General}\ }\textbf {\bibinfo {volume} {33}},\ \bibinfo {pages} {57} (\bibinfo
  {year} {2000})}\BibitemShut {NoStop}%
\bibitem [{\citenamefont {Dal~Negro}\ \emph {et~al.}(2003)\citenamefont
  {Dal~Negro}, \citenamefont {Oton}, \citenamefont {Gaburro}, \citenamefont
  {Pavesi}, \citenamefont {Johnson}, \citenamefont {Lagendijk}, \citenamefont
  {Righini}, \citenamefont {Colocci},\ and\ \citenamefont
  {Wiersma}}]{Negro2003}%
  \BibitemOpen
  \bibfield  {author} {\bibinfo {author} {\bibfnamefont {L.}~\bibnamefont
  {Dal~Negro}}, \bibinfo {author} {\bibfnamefont {C.~J.}\ \bibnamefont {Oton}},
  \bibinfo {author} {\bibfnamefont {Z.}~\bibnamefont {Gaburro}}, \bibinfo
  {author} {\bibfnamefont {L.}~\bibnamefont {Pavesi}}, \bibinfo {author}
  {\bibfnamefont {P.}~\bibnamefont {Johnson}}, \bibinfo {author} {\bibfnamefont
  {A.}~\bibnamefont {Lagendijk}}, \bibinfo {author} {\bibfnamefont
  {R.}~\bibnamefont {Righini}}, \bibinfo {author} {\bibfnamefont
  {M.}~\bibnamefont {Colocci}},\ and\ \bibinfo {author} {\bibfnamefont {D.~S.}\
  \bibnamefont {Wiersma}},\ }\href
  {https://doi.org/10.1103/PhysRevLett.90.055501} {\bibfield  {journal}
  {\bibinfo  {journal} {Phys. Rev. Lett.}\ }\textbf {\bibinfo {volume} {90}},\
  \bibinfo {pages} {055501} (\bibinfo {year} {2003})}\BibitemShut {NoStop}%
\bibitem [{\citenamefont {Fallani}\ \emph {et~al.}(2007)\citenamefont
  {Fallani}, \citenamefont {Lye}, \citenamefont {Guarrera}, \citenamefont
  {Fort},\ and\ \citenamefont {Inguscio}}]{Fallani2007}%
  \BibitemOpen
  \bibfield  {author} {\bibinfo {author} {\bibfnamefont {L.}~\bibnamefont
  {Fallani}}, \bibinfo {author} {\bibfnamefont {J.~E.}\ \bibnamefont {Lye}},
  \bibinfo {author} {\bibfnamefont {V.}~\bibnamefont {Guarrera}}, \bibinfo
  {author} {\bibfnamefont {C.}~\bibnamefont {Fort}},\ and\ \bibinfo {author}
  {\bibfnamefont {M.}~\bibnamefont {Inguscio}},\ }\href
  {https://doi.org/10.1103/PhysRevLett.98.130404} {\bibfield  {journal}
  {\bibinfo  {journal} {Phys. Rev. Lett.}\ }\textbf {\bibinfo {volume} {98}},\
  \bibinfo {pages} {130404} (\bibinfo {year} {2007})}\BibitemShut {NoStop}%
\bibitem [{\citenamefont {Roati}\ \emph {et~al.}(2008)\citenamefont {Roati},
  \citenamefont {D'Errico}, \citenamefont {Fallani}, \citenamefont {Fattori},
  \citenamefont {Fort}, \citenamefont {Zaccanti}, \citenamefont {Modugno},
  \citenamefont {Modugno},\ and\ \citenamefont {Inguscio}}]{Roati2008}%
  \BibitemOpen
  \bibfield  {author} {\bibinfo {author} {\bibfnamefont {G.}~\bibnamefont
  {Roati}}, \bibinfo {author} {\bibfnamefont {C.}~\bibnamefont {D'Errico}},
  \bibinfo {author} {\bibfnamefont {L.}~\bibnamefont {Fallani}}, \bibinfo
  {author} {\bibfnamefont {M.}~\bibnamefont {Fattori}}, \bibinfo {author}
  {\bibfnamefont {C.}~\bibnamefont {Fort}}, \bibinfo {author} {\bibfnamefont
  {M.}~\bibnamefont {Zaccanti}}, \bibinfo {author} {\bibfnamefont
  {G.}~\bibnamefont {Modugno}}, \bibinfo {author} {\bibfnamefont
  {M.}~\bibnamefont {Modugno}},\ and\ \bibinfo {author} {\bibfnamefont
  {M.}~\bibnamefont {Inguscio}},\ }\href {https://doi.org/10.1038/nature07071}
  {\bibfield  {journal} {\bibinfo  {journal} {Nature}\ }\textbf {\bibinfo
  {volume} {453}},\ \bibinfo {pages} {895} (\bibinfo {year}
  {2008})}\BibitemShut {NoStop}%
\bibitem [{\citenamefont {Lahini}\ \emph {et~al.}(2009)\citenamefont {Lahini},
  \citenamefont {Pugatch}, \citenamefont {Pozzi}, \citenamefont {Sorel},
  \citenamefont {Morandotti}, \citenamefont {Davidson},\ and\ \citenamefont
  {Silberberg}}]{Lahini2009}%
  \BibitemOpen
  \bibfield  {author} {\bibinfo {author} {\bibfnamefont {Y.}~\bibnamefont
  {Lahini}}, \bibinfo {author} {\bibfnamefont {R.}~\bibnamefont {Pugatch}},
  \bibinfo {author} {\bibfnamefont {F.}~\bibnamefont {Pozzi}}, \bibinfo
  {author} {\bibfnamefont {M.}~\bibnamefont {Sorel}}, \bibinfo {author}
  {\bibfnamefont {R.}~\bibnamefont {Morandotti}}, \bibinfo {author}
  {\bibfnamefont {N.}~\bibnamefont {Davidson}},\ and\ \bibinfo {author}
  {\bibfnamefont {Y.}~\bibnamefont {Silberberg}},\ }\href
  {https://doi.org/10.1103/PhysRevLett.103.013901} {\bibfield  {journal}
  {\bibinfo  {journal} {Phys. Rev. Lett.}\ }\textbf {\bibinfo {volume} {103}},\
  \bibinfo {pages} {013901} (\bibinfo {year} {2009})}\BibitemShut {NoStop}%
\bibitem [{\citenamefont {Modugno}(2009)}]{Modugno2009}%
  \BibitemOpen
  \bibfield  {author} {\bibinfo {author} {\bibfnamefont {M.}~\bibnamefont
  {Modugno}},\ }\href {https://doi.org/10.1088/1367-2630/11/3/033023}
  {\bibfield  {journal} {\bibinfo  {journal} {New Journal of Physics}\ }\textbf
  {\bibinfo {volume} {11}},\ \bibinfo {pages} {033023} (\bibinfo {year}
  {2009})}\BibitemShut {NoStop}%
\bibitem [{\citenamefont {Kraus}\ \emph {et~al.}(2012)\citenamefont {Kraus},
  \citenamefont {Lahini}, \citenamefont {Ringel}, \citenamefont {Verbin},\ and\
  \citenamefont {Zilberberg}}]{Kraus2012}%
  \BibitemOpen
  \bibfield  {author} {\bibinfo {author} {\bibfnamefont {Y.~E.}\ \bibnamefont
  {Kraus}}, \bibinfo {author} {\bibfnamefont {Y.}~\bibnamefont {Lahini}},
  \bibinfo {author} {\bibfnamefont {Z.}~\bibnamefont {Ringel}}, \bibinfo
  {author} {\bibfnamefont {M.}~\bibnamefont {Verbin}},\ and\ \bibinfo {author}
  {\bibfnamefont {O.}~\bibnamefont {Zilberberg}},\ }\href
  {https://doi.org/10.1103/PhysRevLett.109.106402} {\bibfield  {journal}
  {\bibinfo  {journal} {Phys. Rev. Lett.}\ }\textbf {\bibinfo {volume} {109}},\
  \bibinfo {pages} {106402} (\bibinfo {year} {2012})}\BibitemShut {NoStop}%
\bibitem [{\citenamefont {Segev}\ \emph {et~al.}(2013)\citenamefont {Segev},
  \citenamefont {Silberberg},\ and\ \citenamefont
  {Christodoulides}}]{Segev2013}%
  \BibitemOpen
  \bibfield  {author} {\bibinfo {author} {\bibfnamefont {M.}~\bibnamefont
  {Segev}}, \bibinfo {author} {\bibfnamefont {Y.}~\bibnamefont {Silberberg}},\
  and\ \bibinfo {author} {\bibfnamefont {D.~N.}\ \bibnamefont
  {Christodoulides}},\ }\href@noop {} {\bibfield  {journal} {\bibinfo
  {journal} {Nature Photonics}\ }\textbf {\bibinfo {volume} {7}},\ \bibinfo
  {pages} {197} (\bibinfo {year} {2013})}\BibitemShut {NoStop}%
\bibitem [{\citenamefont {Lellouch}\ and\ \citenamefont
  {Sanchez-Palencia}(2014)}]{Sanchez2014}%
  \BibitemOpen
  \bibfield  {author} {\bibinfo {author} {\bibfnamefont {S.}~\bibnamefont
  {Lellouch}}\ and\ \bibinfo {author} {\bibfnamefont {L.}~\bibnamefont
  {Sanchez-Palencia}},\ }\href {https://doi.org/10.1103/PhysRevA.90.061602}
  {\bibfield  {journal} {\bibinfo  {journal} {Phys. Rev. A}\ }\textbf {\bibinfo
  {volume} {90}},\ \bibinfo {pages} {061602} (\bibinfo {year}
  {2014})}\BibitemShut {NoStop}%
\bibitem [{\citenamefont {Yao}\ \emph {et~al.}(2019)\citenamefont {Yao},
  \citenamefont {Khoudli}, \citenamefont {Bresque},\ and\ \citenamefont
  {Sanchez-Palencia}}]{Sanchez2019}%
  \BibitemOpen
  \bibfield  {author} {\bibinfo {author} {\bibfnamefont {H.}~\bibnamefont
  {Yao}}, \bibinfo {author} {\bibfnamefont {A.}~\bibnamefont {Khoudli}},
  \bibinfo {author} {\bibfnamefont {L.}~\bibnamefont {Bresque}},\ and\ \bibinfo
  {author} {\bibfnamefont {L.}~\bibnamefont {Sanchez-Palencia}},\ }\href
  {https://doi.org/10.1103/PhysRevLett.123.070405} {\bibfield  {journal}
  {\bibinfo  {journal} {Phys. Rev. Lett.}\ }\textbf {\bibinfo {volume} {123}},\
  \bibinfo {pages} {070405} (\bibinfo {year} {2019})}\BibitemShut {NoStop}%
\bibitem [{\citenamefont {Roy}\ \emph {et~al.}(2021)\citenamefont {Roy},
  \citenamefont {Mishra}, \citenamefont {Tanatar},\ and\ \citenamefont
  {Basu}}]{Roy2021}%
  \BibitemOpen
  \bibfield  {author} {\bibinfo {author} {\bibfnamefont {S.}~\bibnamefont
  {Roy}}, \bibinfo {author} {\bibfnamefont {T.}~\bibnamefont {Mishra}},
  \bibinfo {author} {\bibfnamefont {B.}~\bibnamefont {Tanatar}},\ and\ \bibinfo
  {author} {\bibfnamefont {S.}~\bibnamefont {Basu}},\ }\href
  {https://doi.org/10.1103/PhysRevLett.126.106803} {\bibfield  {journal}
  {\bibinfo  {journal} {Phys. Rev. Lett.}\ }\textbf {\bibinfo {volume} {126}},\
  \bibinfo {pages} {106803} (\bibinfo {year} {2021})}\BibitemShut {NoStop}%
\bibitem [{\citenamefont {Ro\'osz}\ \emph {et~al.}(2020)\citenamefont
  {Ro\'osz}, \citenamefont {Zimbor\'as},\ and\ \citenamefont
  {Juh\'asz}}]{Roosz2020}%
  \BibitemOpen
  \bibfield  {author} {\bibinfo {author} {\bibfnamefont {G.~m.~H.}\
  \bibnamefont {Ro\'osz}}, \bibinfo {author} {\bibfnamefont {Z.}~\bibnamefont
  {Zimbor\'as}},\ and\ \bibinfo {author} {\bibfnamefont {R.}~\bibnamefont
  {Juh\'asz}},\ }\href {https://doi.org/10.1103/PhysRevB.102.064204} {\bibfield
   {journal} {\bibinfo  {journal} {Phys. Rev. B}\ }\textbf {\bibinfo {volume}
  {102}},\ \bibinfo {pages} {064204} (\bibinfo {year} {2020})}\BibitemShut
  {NoStop}%
\bibitem [{\citenamefont {Liu}\ \emph {et~al.}(2022)\citenamefont {Liu},
  \citenamefont {Xia}, \citenamefont {Longhi},\ and\ \citenamefont
  {Sanchez-Palencia}}]{Sanchez2022}%
  \BibitemOpen
  \bibfield  {author} {\bibinfo {author} {\bibfnamefont {T.}~\bibnamefont
  {Liu}}, \bibinfo {author} {\bibfnamefont {X.}~\bibnamefont {Xia}}, \bibinfo
  {author} {\bibfnamefont {S.}~\bibnamefont {Longhi}},\ and\ \bibinfo {author}
  {\bibfnamefont {L.}~\bibnamefont {Sanchez-Palencia}},\ }\href
  {https://doi.org/10.21468/SciPostPhys.12.1.027} {\bibfield  {journal}
  {\bibinfo  {journal} {SciPost Phys.}\ }\textbf {\bibinfo {volume} {12}},\
  \bibinfo {pages} {027} (\bibinfo {year} {2022})}\BibitemShut {NoStop}%
\bibitem [{\citenamefont {Yu}\ \emph {et~al.}(2024)\citenamefont {Yu},
  \citenamefont {Bhave}, \citenamefont {Reeve}, \citenamefont {Song},\ and\
  \citenamefont {Schneider}}]{Yu2024}%
  \BibitemOpen
  \bibfield  {author} {\bibinfo {author} {\bibfnamefont {J.-C.}\ \bibnamefont
  {Yu}}, \bibinfo {author} {\bibfnamefont {S.}~\bibnamefont {Bhave}}, \bibinfo
  {author} {\bibfnamefont {L.}~\bibnamefont {Reeve}}, \bibinfo {author}
  {\bibfnamefont {B.}~\bibnamefont {Song}},\ and\ \bibinfo {author}
  {\bibfnamefont {U.}~\bibnamefont {Schneider}},\ }\href
  {https://doi.org/10.1038/s41586-024-07875-2} {\bibfield  {journal} {\bibinfo
  {journal} {Nature}\ }\textbf {\bibinfo {volume} {633}},\ \bibinfo {pages}
  {338} (\bibinfo {year} {2024})}\BibitemShut {NoStop}%
\bibitem [{Roo()}]{Roosz2014}%
  \BibitemOpen
  \href@noop {} {\ }\BibitemShut {NoStop}%
\bibitem [{\citenamefont {Sinha}\ \emph {et~al.}(2019)\citenamefont {Sinha},
  \citenamefont {Rams},\ and\ \citenamefont {Dziarmaga}}]{Dziarmaga2019}%
  \BibitemOpen
  \bibfield  {author} {\bibinfo {author} {\bibfnamefont {A.}~\bibnamefont
  {Sinha}}, \bibinfo {author} {\bibfnamefont {M.~M.}\ \bibnamefont {Rams}},\
  and\ \bibinfo {author} {\bibfnamefont {J.}~\bibnamefont {Dziarmaga}},\ }\href
  {https://doi.org/10.1103/PhysRevB.99.094203} {\bibfield  {journal} {\bibinfo
  {journal} {Phys. Rev. B}\ }\textbf {\bibinfo {volume} {99}},\ \bibinfo
  {pages} {094203} (\bibinfo {year} {2019})}\BibitemShut {NoStop}%
\bibitem [{\citenamefont {Ro\'osz}(2024)}]{Roosz2024}%
  \BibitemOpen
  \bibfield  {author} {\bibinfo {author} {\bibfnamefont {G.~m.~H.}\
  \bibnamefont {Ro\'osz}},\ }\href
  {https://doi.org/10.1103/PhysRevB.109.064204} {\bibfield  {journal} {\bibinfo
   {journal} {Phys. Rev. B}\ }\textbf {\bibinfo {volume} {109}},\ \bibinfo
  {pages} {064204} (\bibinfo {year} {2024})}\BibitemShut {NoStop}%
\bibitem [{\citenamefont {Cai}\ \emph {et~al.}(2013)\citenamefont {Cai},
  \citenamefont {Lang}, \citenamefont {Chen},\ and\ \citenamefont
  {Wang}}]{Cai2013}%
  \BibitemOpen
  \bibfield  {author} {\bibinfo {author} {\bibfnamefont {X.}~\bibnamefont
  {Cai}}, \bibinfo {author} {\bibfnamefont {L.-J.}\ \bibnamefont {Lang}},
  \bibinfo {author} {\bibfnamefont {S.}~\bibnamefont {Chen}},\ and\ \bibinfo
  {author} {\bibfnamefont {Y.}~\bibnamefont {Wang}},\ }\href
  {https://doi.org/10.1103/PhysRevLett.110.176403} {\bibfield  {journal}
  {\bibinfo  {journal} {Phys. Rev. Lett.}\ }\textbf {\bibinfo {volume} {110}},\
  \bibinfo {pages} {176403} (\bibinfo {year} {2013})}\BibitemShut {NoStop}%
\bibitem [{\citenamefont {Wang}\ \emph {et~al.}(2016)\citenamefont {Wang},
  \citenamefont {Liu}, \citenamefont {Xianlong},\ and\ \citenamefont
  {Hu}}]{Wang2016}%
  \BibitemOpen
  \bibfield  {author} {\bibinfo {author} {\bibfnamefont {J.}~\bibnamefont
  {Wang}}, \bibinfo {author} {\bibfnamefont {X.-J.}\ \bibnamefont {Liu}},
  \bibinfo {author} {\bibfnamefont {G.}~\bibnamefont {Xianlong}},\ and\
  \bibinfo {author} {\bibfnamefont {H.}~\bibnamefont {Hu}},\ }\href
  {https://doi.org/10.1103/PhysRevB.93.104504} {\bibfield  {journal} {\bibinfo
  {journal} {Phys. Rev. B}\ }\textbf {\bibinfo {volume} {93}},\ \bibinfo
  {pages} {104504} (\bibinfo {year} {2016})}\BibitemShut {NoStop}%
\bibitem [{\citenamefont {Zeng}\ \emph {et~al.}(2016)\citenamefont {Zeng},
  \citenamefont {Chen},\ and\ \citenamefont {L\"u}}]{Chen2016}%
  \BibitemOpen
  \bibfield  {author} {\bibinfo {author} {\bibfnamefont {Q.-B.}\ \bibnamefont
  {Zeng}}, \bibinfo {author} {\bibfnamefont {S.}~\bibnamefont {Chen}},\ and\
  \bibinfo {author} {\bibfnamefont {R.}~\bibnamefont {L\"u}},\ }\href
  {https://doi.org/10.1103/PhysRevB.94.125408} {\bibfield  {journal} {\bibinfo
  {journal} {Phys. Rev. B}\ }\textbf {\bibinfo {volume} {94}},\ \bibinfo
  {pages} {125408} (\bibinfo {year} {2016})}\BibitemShut {NoStop}%
\bibitem [{\citenamefont {Crowley}\ \emph {et~al.}(2018)\citenamefont
  {Crowley}, \citenamefont {Chandran},\ and\ \citenamefont
  {Laumann}}]{Chandran2018}%
  \BibitemOpen
  \bibfield  {author} {\bibinfo {author} {\bibfnamefont {P.~J.~D.}\
  \bibnamefont {Crowley}}, \bibinfo {author} {\bibfnamefont {A.}~\bibnamefont
  {Chandran}},\ and\ \bibinfo {author} {\bibfnamefont {C.~R.}\ \bibnamefont
  {Laumann}},\ }\href {https://doi.org/10.1103/PhysRevLett.120.175702}
  {\bibfield  {journal} {\bibinfo  {journal} {Phys. Rev. Lett.}\ }\textbf
  {\bibinfo {volume} {120}},\ \bibinfo {pages} {175702} (\bibinfo {year}
  {2018})}\BibitemShut {NoStop}%
\bibitem [{\citenamefont {Tong}\ \emph {et~al.}(2021)\citenamefont {Tong},
  \citenamefont {Meng}, \citenamefont {Jiang}, \citenamefont {Lee},
  \citenamefont {Neto},\ and\ \citenamefont {Xianlong}}]{Gao2021}%
  \BibitemOpen
  \bibfield  {author} {\bibinfo {author} {\bibfnamefont {X.}~\bibnamefont
  {Tong}}, \bibinfo {author} {\bibfnamefont {Y.-M.}\ \bibnamefont {Meng}},
  \bibinfo {author} {\bibfnamefont {X.}~\bibnamefont {Jiang}}, \bibinfo
  {author} {\bibfnamefont {C.}~\bibnamefont {Lee}}, \bibinfo {author}
  {\bibfnamefont {G.~D. d.~M.}\ \bibnamefont {Neto}},\ and\ \bibinfo {author}
  {\bibfnamefont {G.}~\bibnamefont {Xianlong}},\ }\href
  {https://doi.org/10.1103/PhysRevB.103.104202} {\bibfield  {journal} {\bibinfo
   {journal} {Phys. Rev. B}\ }\textbf {\bibinfo {volume} {103}},\ \bibinfo
  {pages} {104202} (\bibinfo {year} {2021})}\BibitemShut {NoStop}%
\bibitem [{\citenamefont {Cheng}\ \emph {et~al.}(2023)\citenamefont {Cheng},
  \citenamefont {Asgari},\ and\ \citenamefont {Xianlong}}]{Gao2023}%
  \BibitemOpen
  \bibfield  {author} {\bibinfo {author} {\bibfnamefont {S.}~\bibnamefont
  {Cheng}}, \bibinfo {author} {\bibfnamefont {R.}~\bibnamefont {Asgari}},\ and\
  \bibinfo {author} {\bibfnamefont {G.}~\bibnamefont {Xianlong}},\ }\href
  {https://doi.org/10.1103/PhysRevB.108.024204} {\bibfield  {journal} {\bibinfo
   {journal} {Phys. Rev. B}\ }\textbf {\bibinfo {volume} {108}},\ \bibinfo
  {pages} {024204} (\bibinfo {year} {2023})}\BibitemShut {NoStop}%
\bibitem [{\citenamefont {Lv}\ \emph {et~al.}(2022{\natexlab{a}})\citenamefont
  {Lv}, \citenamefont {Yi}, \citenamefont {Li}, \citenamefont {Sun},\ and\
  \citenamefont {You}}]{You2022-PRA}%
  \BibitemOpen
  \bibfield  {author} {\bibinfo {author} {\bibfnamefont {T.}~\bibnamefont
  {Lv}}, \bibinfo {author} {\bibfnamefont {T.-C.}\ \bibnamefont {Yi}}, \bibinfo
  {author} {\bibfnamefont {L.}~\bibnamefont {Li}}, \bibinfo {author}
  {\bibfnamefont {G.}~\bibnamefont {Sun}},\ and\ \bibinfo {author}
  {\bibfnamefont {W.-L.}\ \bibnamefont {You}},\ }\href
  {https://doi.org/10.1103/PhysRevA.105.013315} {\bibfield  {journal} {\bibinfo
   {journal} {Phys. Rev. A}\ }\textbf {\bibinfo {volume} {105}},\ \bibinfo
  {pages} {013315} (\bibinfo {year} {2022}{\natexlab{a}})}\BibitemShut
  {NoStop}%
\bibitem [{\citenamefont {Lv}\ \emph {et~al.}(2022{\natexlab{b}})\citenamefont
  {Lv}, \citenamefont {Liu}, \citenamefont {Yi}, \citenamefont {Li},
  \citenamefont {Liu},\ and\ \citenamefont {You}}]{You2022-PRB}%
  \BibitemOpen
  \bibfield  {author} {\bibinfo {author} {\bibfnamefont {T.}~\bibnamefont
  {Lv}}, \bibinfo {author} {\bibfnamefont {Y.-B.}\ \bibnamefont {Liu}},
  \bibinfo {author} {\bibfnamefont {T.-C.}\ \bibnamefont {Yi}}, \bibinfo
  {author} {\bibfnamefont {L.}~\bibnamefont {Li}}, \bibinfo {author}
  {\bibfnamefont {M.}~\bibnamefont {Liu}},\ and\ \bibinfo {author}
  {\bibfnamefont {W.-L.}\ \bibnamefont {You}},\ }\href
  {https://doi.org/10.1103/PhysRevB.106.144205} {\bibfield  {journal} {\bibinfo
   {journal} {Phys. Rev. B}\ }\textbf {\bibinfo {volume} {106}},\ \bibinfo
  {pages} {144205} (\bibinfo {year} {2022}{\natexlab{b}})}\BibitemShut
  {NoStop}%
\bibitem [{\citenamefont {Roy}\ \emph {et~al.}(2024)\citenamefont {Roy},
  \citenamefont {Roy},\ and\ \citenamefont {Basu}}]{Roy2024}%
  \BibitemOpen
  \bibfield  {author} {\bibinfo {author} {\bibfnamefont {K.}~\bibnamefont
  {Roy}}, \bibinfo {author} {\bibfnamefont {S.}~\bibnamefont {Roy}},\ and\
  \bibinfo {author} {\bibfnamefont {S.}~\bibnamefont {Basu}},\ }\href@noop {}
  {\bibfield  {journal} {\bibinfo  {journal} {Scientific Reports}\ }\textbf
  {\bibinfo {volume} {14}},\ \bibinfo {pages} {20603} (\bibinfo {year}
  {2024})}\BibitemShut {NoStop}%
\bibitem [{\citenamefont {Caneva}\ \emph {et~al.}(2007)\citenamefont {Caneva},
  \citenamefont {Fazio},\ and\ \citenamefont {Santoro}}]{Caneva2007}%
  \BibitemOpen
  \bibfield  {author} {\bibinfo {author} {\bibfnamefont {T.}~\bibnamefont
  {Caneva}}, \bibinfo {author} {\bibfnamefont {R.}~\bibnamefont {Fazio}},\ and\
  \bibinfo {author} {\bibfnamefont {G.~E.}\ \bibnamefont {Santoro}},\ }\href
  {https://doi.org/10.1103/PhysRevB.76.144427} {\bibfield  {journal} {\bibinfo
  {journal} {Phys. Rev. B}\ }\textbf {\bibinfo {volume} {76}},\ \bibinfo
  {pages} {144427} (\bibinfo {year} {2007})}\BibitemShut {NoStop}%
\bibitem [{\citenamefont {Ino}\ and\ \citenamefont {Kohmoto}(2006)}]{Ino2006}%
  \BibitemOpen
  \bibfield  {author} {\bibinfo {author} {\bibfnamefont {K.}~\bibnamefont
  {Ino}}\ and\ \bibinfo {author} {\bibfnamefont {M.}~\bibnamefont {Kohmoto}},\
  }\href {https://doi.org/10.1103/PhysRevB.73.205111} {\bibfield  {journal}
  {\bibinfo  {journal} {Phys. Rev. B}\ }\textbf {\bibinfo {volume} {73}},\
  \bibinfo {pages} {205111} (\bibinfo {year} {2006})}\BibitemShut {NoStop}%
\bibitem [{\citenamefont {Dziarmaga}(2006)}]{Dziarmaga2006}%
  \BibitemOpen
  \bibfield  {author} {\bibinfo {author} {\bibfnamefont {J.}~\bibnamefont
  {Dziarmaga}},\ }\href {https://doi.org/10.1103/PhysRevB.74.064416} {\bibfield
   {journal} {\bibinfo  {journal} {Phys. Rev. B}\ }\textbf {\bibinfo {volume}
  {74}},\ \bibinfo {pages} {064416} (\bibinfo {year} {2006})}\BibitemShut
  {NoStop}%
\bibitem [{\citenamefont {Dziarmaga}(2010)}]{Dziarmaga2010}%
  \BibitemOpen
  \bibfield  {author} {\bibinfo {author} {\bibfnamefont {J.}~\bibnamefont
  {Dziarmaga}},\ }\href {https://doi.org/10.1080/00018732.2010.514702}
  {\bibfield  {journal} {\bibinfo  {journal} {Advances in Physics}\ }\textbf
  {\bibinfo {volume} {59}},\ \bibinfo {pages} {1063} (\bibinfo {year}
  {2010})}\BibitemShut {NoStop}%
\bibitem [{\citenamefont {Francuz}\ \emph {et~al.}(2016)\citenamefont
  {Francuz}, \citenamefont {Dziarmaga}, \citenamefont {Gardas},\ and\
  \citenamefont {Zurek}}]{Zurek2016}%
  \BibitemOpen
  \bibfield  {author} {\bibinfo {author} {\bibfnamefont {A.}~\bibnamefont
  {Francuz}}, \bibinfo {author} {\bibfnamefont {J.}~\bibnamefont {Dziarmaga}},
  \bibinfo {author} {\bibfnamefont {B.}~\bibnamefont {Gardas}},\ and\ \bibinfo
  {author} {\bibfnamefont {W.~H.}\ \bibnamefont {Zurek}},\ }\href
  {https://doi.org/10.1103/PhysRevB.93.075134} {\bibfield  {journal} {\bibinfo
  {journal} {Phys. Rev. B}\ }\textbf {\bibinfo {volume} {93}},\ \bibinfo
  {pages} {075134} (\bibinfo {year} {2016})}\BibitemShut {NoStop}%
\bibitem [{\citenamefont {Deng}\ \emph {et~al.}(2009)\citenamefont {Deng},
  \citenamefont {Ortiz},\ and\ \citenamefont {Viola}}]{Deng2008}%
  \BibitemOpen
  \bibfield  {author} {\bibinfo {author} {\bibfnamefont {S.}~\bibnamefont
  {Deng}}, \bibinfo {author} {\bibfnamefont {G.}~\bibnamefont {Ortiz}},\ and\
  \bibinfo {author} {\bibfnamefont {L.}~\bibnamefont {Viola}},\ }\href
  {https://doi.org/10.1209/0295-5075/84/67008} {\bibfield  {journal} {\bibinfo
  {journal} {Europhysics Letters}\ }\textbf {\bibinfo {volume} {84}},\ \bibinfo
  {pages} {67008} (\bibinfo {year} {2009})}\BibitemShut {NoStop}%
\bibitem [{\citenamefont {Kolodrubetz}\ \emph {et~al.}(2012)\citenamefont
  {Kolodrubetz}, \citenamefont {Clark},\ and\ \citenamefont
  {Huse}}]{David2012}%
  \BibitemOpen
  \bibfield  {author} {\bibinfo {author} {\bibfnamefont {M.}~\bibnamefont
  {Kolodrubetz}}, \bibinfo {author} {\bibfnamefont {B.~K.}\ \bibnamefont
  {Clark}},\ and\ \bibinfo {author} {\bibfnamefont {D.~A.}\ \bibnamefont
  {Huse}},\ }\href {https://doi.org/10.1103/PhysRevLett.109.015701} {\bibfield
  {journal} {\bibinfo  {journal} {Phys. Rev. Lett.}\ }\textbf {\bibinfo
  {volume} {109}},\ \bibinfo {pages} {015701} (\bibinfo {year}
  {2012})}\BibitemShut {NoStop}%
\bibitem [{\citenamefont {Dziarmaga}\ \emph {et~al.}(2022)\citenamefont
  {Dziarmaga}, \citenamefont {Rams},\ and\ \citenamefont {Zurek}}]{Zurek2022}%
  \BibitemOpen
  \bibfield  {author} {\bibinfo {author} {\bibfnamefont {J.}~\bibnamefont
  {Dziarmaga}}, \bibinfo {author} {\bibfnamefont {M.~M.}\ \bibnamefont
  {Rams}},\ and\ \bibinfo {author} {\bibfnamefont {W.~H.}\ \bibnamefont
  {Zurek}},\ }\href {https://doi.org/10.1103/PhysRevLett.129.260407} {\bibfield
   {journal} {\bibinfo  {journal} {Phys. Rev. Lett.}\ }\textbf {\bibinfo
  {volume} {129}},\ \bibinfo {pages} {260407} (\bibinfo {year}
  {2022})}\BibitemShut {NoStop}%
\end{thebibliography}%

 \end{document}